\documentclass[reprint,superscriptaddress,amsmath,amssymb,
 aps,showkeys]{revtex4-2}
\usepackage{graphicx}
\usepackage{dcolumn}
\usepackage{bm}
\usepackage{multirow}
\usepackage{booktabs}
\usepackage{float}
\usepackage{placeins}
\usepackage{xcolor}
\usepackage[colorlinks=true,linkcolor=blue,citecolor=blue,urlcolor=blue]{hyperref}
\usepackage{miller}
\renewcommand{\figurename}{Figure}
\renewcommand{\tablename}{Table}

\begin{document}

\preprint{APS/123-QED}

\title{Beyond Fundamental Building Blocks: Plasticity in Structurally Complex Crystals}

\author{Tobias Stollenwerk}
\email[]{stollenwerk@imm.rwth-aachen.de}
\affiliation{Institute of Physical Metallurgy and Materials Physics, RWTH Aachen University, 52056 Aachen, Germany}
    
\author{Pia Carlotta Huckfeldt}
\affiliation{Institute of Physical Metallurgy and Materials Physics, RWTH Aachen University, 52056 Aachen, Germany}

\author{Nisa Zakia Zahra Ulumuddin}
\affiliation{Institute of Physical Metallurgy and Materials Physics, RWTH Aachen University, 52056 Aachen, Germany}

\author{Malik Schneider}
\affiliation{Institute of Physical Metallurgy and Materials Physics, RWTH Aachen University, 52056 Aachen, Germany}

\author{Zhuocheng Xie}
\email[]{xie@imm.rwth-aachen.de}
\affiliation{Institute of Physical Metallurgy and Materials Physics, RWTH Aachen University, 52056 Aachen, Germany}

\author{Sandra Korte-Kerzel}
\email[]{korte-kerzel@imm.rwth-aachen.de}
\affiliation{Institute of Physical Metallurgy and Materials Physics, RWTH Aachen University, 52056 Aachen, Germany}

\begin{abstract}
Intermetallics, which encompass a wide range of compounds, often exhibit similar or closely related crystal structures, resulting in various intermetallic systems with structurally derivative phases. This study examines the hypothesis that deformation behavior can be transferred from fundamental building blocks to structurally related phases using the binary samarium-cobalt system. We investigate SmCo$_2$ and SmCo$_5$ as fundamental building blocks and compare them to the structurally related SmCo$_3$ and Sm$_2$Co$_{17}$ phases. Nanoindentation and micropillar compression tests were performed to characterize the primary slip systems, complemented by generalized stacking fault energy calculations via atomic-scale modeling. Our results show that while elastic properties of the structurally complex phases follow a rule of mixtures, their plastic deformation mechanisms are more intricate, influenced by the stacking and bonding nature within the crystal's building blocks. These findings underscore the importance of local bonding environments in predicting the mechanical behavior of structurally related intermetallics, providing crucial insights for the development of high-performance intermetallic materials.
\end{abstract}


\maketitle

\section{Introduction}\label{introduction}

Intermetallic phases, particularly topologically close-packed phases, often exhibit similar or closely related crystal structures, which results in a plethora of intermetallic systems containing different phases that are structural derivatives of each other \cite{sinha1972,Sauthoff1995}. It is therefore often hypothesized that the material properties of these phases are significantly related. By investigating the relationship of the material properties among the crystal's fundamental building blocks within one intermetallic system, it may be possible to extrapolate findings to other, more complex, structurally related phases  \cite{liu1984ductile,Howie2017,kolli2020discovering,Luo2023}. 

Such a relationship was previously demonstrated in the deformation behavior of a topologically close-packed intermetallic system, where the C15 Laves phase (space group Fd$\bar{3}$m, MgCu$_2$ structure type) and the Zr$_4$Al$_3$ phase (space group P6/mmm, Zr$_4$Al$_3$ structure type) serve as the fundamental building blocks comprising the $\mu$ phase (space group R$\bar{3}$m H, W$_6$Fe$_7$ structure type). The elastic moduli of the $\mu$ phase follow the rule of mixtures, lying between the values of the Laves-phase and Zr$_4$Al$_3$-phase building blocks \cite{Luo2023}, where the Laves phase is more compliant than the Zr$_4$Al$_3$ phase. Plastic deformation in the $\mu$ phases occurs within the compliant Laves-phase building blocks \cite{Schroders2018,Schroders2019,Luo2023,luo2023plasticity} in the form of synchro-shear or crystallographic slip, which are typical deformation mechanisms in Laves phases, as demonstrated experimentally and through simulations \cite{Kronberg1957,Hazzledine1992,Chisholm2005,vedmedenko2008,guenole2019,Xie2023,Xie2023a,Wang2024}. Whether this relationship can be generalized to other intermetallic systems remains unknown.

\begin{figure*}[hbt!]
\centering
\includegraphics[width=0.75\textwidth]{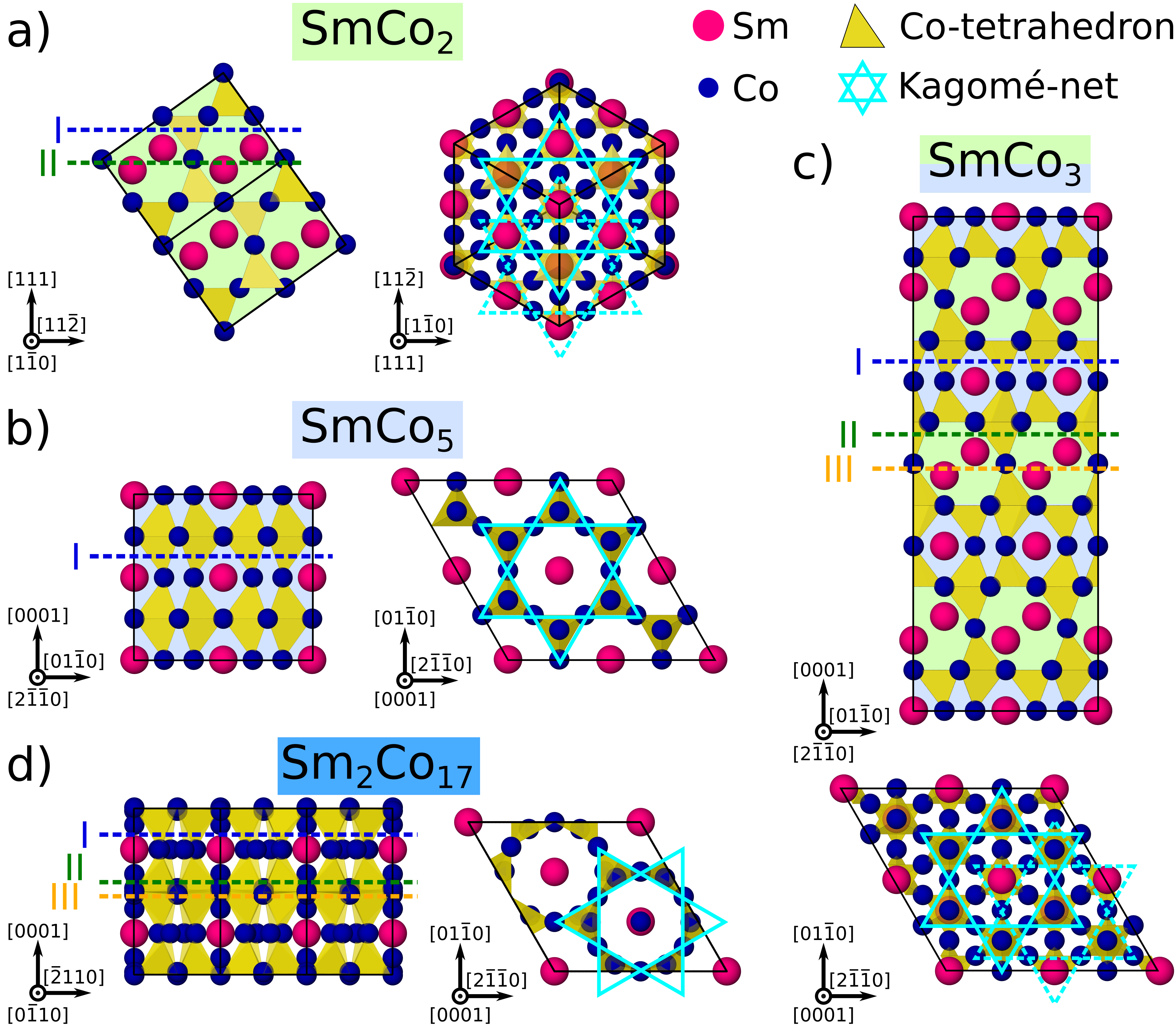}
\caption{Schematics of the crystal structures for the binary Sm-Co systems. a) SmCo$_{2}$ in space group Fd$\bar{3}$m and the MgCu$_2$ structure type, classified as a C15 Laves phase. Co-tetrahedra (colored in yellow) are arranged in an alternating stacking sequence and connected via the corners. b) SmCo$_{5}$ in space group P6/mmm and the CaCu$_5$ structure type. The Co-tetrahedra are connected along a plane to form trigonal bipyramids, which are connected via their corners. c) SmCo$_{3}$ in space group R$\bar{3}$m H and the PuNi$_3$ structure type, comprising an alternating stacking of the MgCu$_2$ and CaCu$_5$ structure types. d) Sm$_{2}$Co$_{17}$ in space group P63/mmc and the Th$_2$Ni$_{17}$ structure type. The Co-tetrahedra are arranged in trigonal bipyramids. Its crystal structure can also be described as a variation of SmCo$_5$, where a third of the Sm atoms are replaced with two Co atoms. In these four Sm-Co crystal structures, Co Kagomé-nets (colored in cyan) that spread through the crystal structure perpendicular to the $\langle1\,1\,1\rangle$ or $\langle0\,0\,0\,1\rangle$ directions. Large (colored in red) and small (colored in blue) atoms are Sm and Co atoms, respectively. The dotted lines indicate the $(1\,1\,1)$ and basal planes between different atomic layers.}
\label{fig1}
\end{figure*}

Here, we therefore systematically investigate a single binary system which offers several structurally related phases built upon few fundamental building blocks: Sm-Co. There are primarily two phases, SmCo$_2$ and SmCo$_5$, whose structural motifs can be found in other more structurally complex phases of the system \cite{Buschow1968,Westbrook1995}. In this sense, these two phases act as fundamental building blocks for phases such as SmCo$_3$ \cite{Lewy1965, Buschow1969}, Sm$_2$Co$_7$ \cite{Campos2000}, Sm$_5$Co$_{19}$ \cite{Takeda1982,Derkaoui1989}, and Sm$_2$Co$_{17}$ \cite{Buschow1966,Campos2000}, which exhibit layered crystal structures incorporating these building blocks in different stacking variations. Figure \ref{fig1} schematically depicts the structural relationship between the different phases, focusing on the stacking of the various building block layers and highlighting the interfacing Co Kagomé-nets along the $(1\,1\,1)$ plane and the basal plane. The intermetallic Sm-Co system is well characterized with regard to its magnetic properties, especially in high-temperature permanent magnet applications, due to its exceptional coercivity and thermal stability \cite{Nesbitt1962,Ray1992,Chen1998,Goll2000,Gutfleisch2006}. However, the deformation mechanisms of the Sm-Co intermetallic phases remain largely unexplored. The foundations for this endeavor were laid in one of our previous works \cite{Stollenwerk2024}, where we examined the plasticity in the primary SmCo$_5$ phase. Contrary to an earlier study suggesting that amorphous shear bands were considered to mediate plasticity \cite{Luo2019,Luo2020}, we found that plastic deformation in SmCo$_5$ occurs via dislocation motion along distinct crystallographic planes, including a pyramidal $\{2\,\bar{1}\,\bar{1}\, \bar{1}\}$ $\langle2\,1\,1\,\bar{6} \rangle$ slip system and a basal $(0\,0\,0\,1)$ $[2\,\bar{1}\, \bar{1}\,0]$ slip system.

This study aims to examine the hypothesis that mechanical properties and deformation mechanisms can be transferred from fundamental building blocks to structurally related intermetallic phases. We compared the deformation behavior of the primary SmCo$_2$ and SmCo$_5$ phases with the structurally complex SmCo$_3$ and Sm$_2$Co$_{17}$ phases, focusing particularly on investigating and characterizing basal and $(1\,1\,1)$ slip, as these correspond to the planes along which the building blocks are stacking to form the larger unit cells. To study the deformation behavior of the four Sm-Co phases, we utilized a combination of experimental studies and atomic-scale modeling. We conducted nanoindentation tests to establish an overview of the global mechanical properties, like hardness and indentation modulus, and to gain initial insights into potential slip systems by analyzing the slip traces around the indents. These initial observations were then scrutinized via micropillar compression testing to associate the activated slip systems with their specific critical resolved shear stresses (CRSS). These experimental findings were combined with modelling. This included atomistic simulations to calculate the generalized stacking fault energy (GSFE) and consequent changes in the barriers to dislocation motion, as well as density functional theory (DFT) calculations to reveal intricate changes in bonding characteristics for the building blocks as they are combined to form larger unit cells.

\section{Results}\label{results}

\subsection{Nanoindentation}

We performed nanoindentation to investigate the orientation dependence of hardness and indentation modulus, and induce slip trace formation for an initial assessment of the relative activity of individual slip planes. Different orientations of the measured grains were chosen for each phase, as illustrated in the inverse pole figures (IPF) in Figure S4 in the SI. The aim was to cover a significant portion of the standard triangle to activate various slip systems, particularly slip on the $\{1\,1\,1\}$ and basal planes. The summary of the indentation tests and results for each of the phases is listed in Table \ref{table:1}.

\begin{figure*}[p!]
\centering
\includegraphics[width=0.75\textwidth]{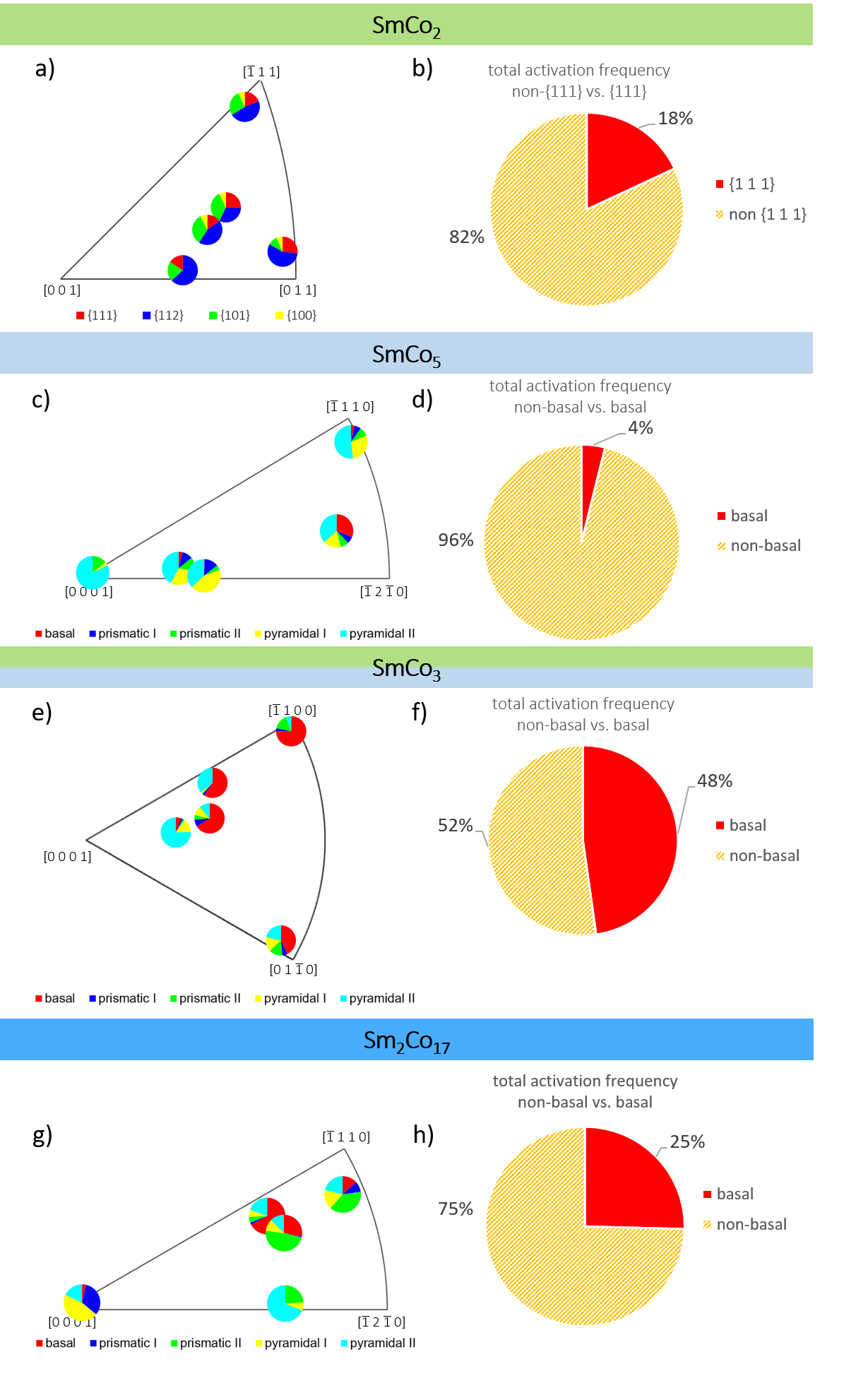}
\caption{Slip trace analysis for the four Sm-Co phases after nanoindentation tests: a) Slip trace distribution in SmCo$_{2}$ with regard to grain orientation; b) Contribution of basal slip in SmCo$_{2}$; c) Slip trace distribution in SmCo$_{5}$ with regard to grain orientation; d) Contribution of basal slip in SmCo$_{5}$; e) Slip trace distribution in SmCo$_{3}$ with regard to grain orientation; f) Contribution of basal slip in SmCo$_{3}$; g) Slip trace distribution in Sm$_{2}$Co$_{17}$ with regard to grain orientation; h) Contribution of basal slip in Sm$_{2}$Co$_{17}$.}
\label{fig2}
\end{figure*}

Across the grains indented in the SmCo$_2$, we found no significant orientation dependence of hardness and indentation modulus, for which the average values were determined as 6.3 ± 0.2 GPa and 72 ± 5 GPa, respectively. Around these indentations, the slip traces  showed a distribution of 18\% attributed to $\{1\,1\,1\}$ and 82\% to non-$\{1\,1\,1\}$ planes (including \hkl{1 1 1}, \hkl{1 1 2}, \hkl{1 0 1}, and \hkl{100} planes), see Figure \ref{fig2}a-b.
There were remaining slip traces that could not be identified as any of these low-index planes. These correspond to higher index planes, most likely of $\{1\,1\,n\}$ type \cite{Freund2024}.  These planes are closely spaced and can therefore not be distinguished by surface slip trace analysis, leading to a strong distortion of the data through double counting. The fraction of non-\hkl{1 1 1} planes is therefore a slight underestimate. 

The slip trace analysis for indents in the SmCo$_5$ phase indicated that pyramidal I and II slip were the major contributors to deformation during nanoindentation. Basal slip was primarily observed at locations in the standard triangle with a high Schmid factor for basal slip (Figure \ref{fig2}c), contributing to only 4\% of the total slip traces (Figure \ref{fig2}d). 

Except for the orientation closest to the basal orientation, all grains in SmCo$_{3}$ show a significant amount of basal slip trace around the indents (Figure \ref{fig2}e). Hence, about 48\% of the observed slip traces were indexed as basal (Figure \ref{fig2}f). We noted in particular, that the high fraction of basal traces in those grains where the plane is oriented approximately parallel to the indentation axis, is associated with a faint appearance of the surface traces.

In Sm$_2$Co$_{17}$, the slip trace analysis showed a mixed distribution of slip traces, with the quarter of traces corresponding to basal plane activation observed predominately in orientations where the basal plane has an intermediate inclination angle (Figure \ref{fig2}g-h).

\begin{table}[hpt!]
\centering
\small

\caption{Summary of the indentation tests including hardness and indentation modulus values for the four Sm-Co phases.}
\label{table:1}
\begin{tabular}{>{\centering\arraybackslash}m{0.3\linewidth}>{\centering\arraybackslash}m{0.15\linewidth}>{\centering\arraybackslash}m{0.15\linewidth}>{\centering\arraybackslash}m{0.15\linewidth}>{\centering\arraybackslash}m{0.15\linewidth}}
\addlinespace[0.1cm]
\hline\hline      
 & SmCo$_2$& SmCo$_5$& SmCo$_3$& Sm$_2$Co$_{17}$\\ \hline
\addlinespace[0.1cm]
No. orientations & 5 & 5 & 5 & 5 \\ \hline
\addlinespace[0.1cm]
No. indents & 65 & 234 & 118 & 257 \\ \hline
\addlinespace[0.1cm]
Aver. hardness [GPa]& 6.3$\pm$0.2& 8.7$\pm$0.2& 7.8$\pm$0.4& 8.5$\pm$0.5\\ \hline
\addlinespace[0.1cm]
Aver. ind. modulus [GPa]& 72$\pm$5& 131$\pm$4& 97$\pm$4& 182$\pm$3\\ \hline\hline

\end{tabular}

\end{table}

\subsection{Micropillar Compression}

 We performed compression tests on a total of 82 Sm-Co micropillars (and an additional 30 from a previous study \cite{Stollenwerk2024}) in order to establish quantitative measurements of CRSS (see Table \ref{table:2}), particularly of the plastic events on the basal and $\{1\,1\,1\}$ planes in addition to the assessment of their relative activation compared with non-basal or non-$\{1\,1\,1\}$ slip and measurements of hardness by nanoindentation. The grain orientations for each phase are given in Figure S5 in the SI. The grains of none of the phases were large enough to allow both indentations and compression tests to be performed on the same grain. Consequently, the orientations of the micropillars differ from those in nanoindentation measurements and were chosen to target specific slip systems from the indentation slip trace analysis by maximizing their (uniaxial) Schmid factor ($m$). The following paragraphs describe the determined slip systems for each phase.

\begin{table*}[hpt!]
\centering
\small

\caption{Summary of the micropillar compression tests including activated slip systems for the four Sm-Co phases.}
\label{table:2}
\begin{tabular}{>{\centering\arraybackslash}m{0.2\linewidth}>{\centering\arraybackslash}m{0.12\linewidth}>{\centering\arraybackslash}m{0.12\linewidth}>{\centering\arraybackslash}m{0.12\linewidth}>{\centering\arraybackslash}m{0.12\linewidth}} \hline\hline   \addlinespace[0.1cm]
 & SmCo$_2$& SmCo$_5$& SmCo$_3$& Sm$_2$Co$_{17}$\\ \hline 
\addlinespace[0.1cm]
No. orientations& 5 & 2 & 10 & 3 \\ \hline 
\addlinespace[0.1cm]
No. pillars& 20 & 30 & 32 & 30 \\ \hline
\addlinespace[0.1cm]
Activated slip systems & $(1\,\bar{1}\,1)[\bar{1}\,0\,1]$ $(1\,1\,2)[2\,0\,\bar{1}]$ $(1\,\bar{6}\,\bar{1)}[\bar{1}\,0\,\bar{1}]$& $(2\,\bar{1}\,\bar{1}\,\bar{1})[\bar{2}\,1\,1\,\bar{6}]$ $(0\,0\,0\,1)[2\,\bar{1}\,\bar{1}\,0]$  $(0\,\bar{1}\,1\,1)[2\,\bar{1}\,\bar{1}\,0]$& $\{0\,1\,\bar{1}\,0\}\langle2\,\bar{1}\,\bar{1}\,0\rangle$ $(0\,0\,0\,1)[2\,\bar{1}\,\bar{1}\,0]$& $(0\,0\,0\,1)[2\,\bar{1}\,\bar{1}\,0]$ $(\bar{1}\,\bar{1}\,2\,0)[1\,\bar{1}\,0\,0]$  $\{\bar{1}\,1\,0\,1\}\langle2\,\bar{1}\,\bar{1}\,0\rangle$\\ \hline\hline

\end{tabular}

\end{table*}

\begin{figure*}[t!]
\centering
\includegraphics[width=0.875\textwidth]{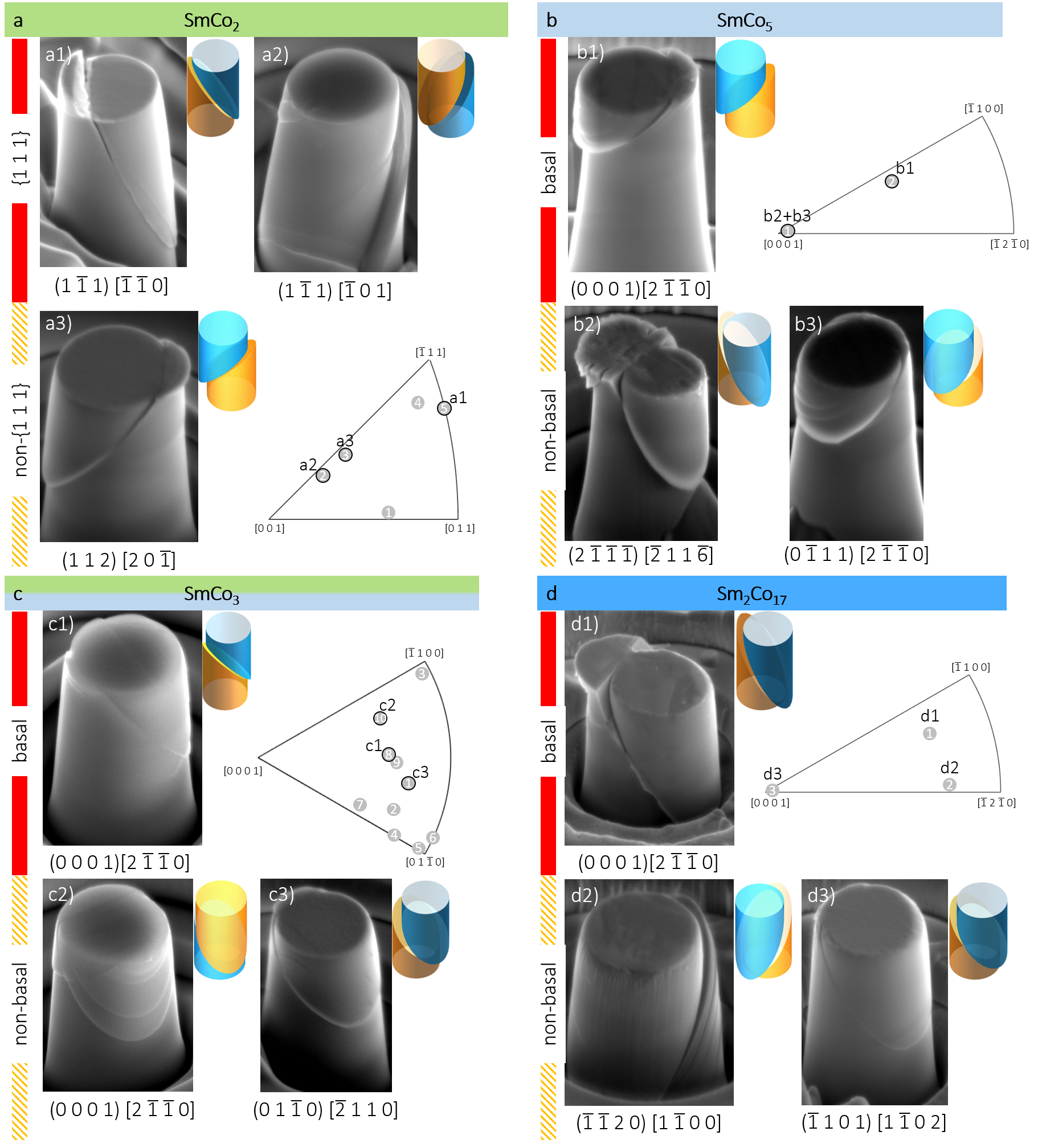}
\caption{Deformed micropillars of the a) SmCo$_2$, b) SmCo$_5$, SmCo$_3$, and Sm$_2$Co$_{17}$ phases after compression tests. a1) Activation of the $(1 \,\bar{1}\,1)[\bar{1}\,\bar{1}\,0]$ slip system. a2-a4) Other slip systems activated in the SmCo$_2$ phase. b1) Activation of the $(0\,0\,0\,1)[2\,\bar{1}\,\bar{1}\,0]$ slip system. b2-b4) Other slip systems were activated in the SmCo$_5$ phase. c1) Activation of the $(0\,0\,0\,1)[2\,\bar{1}\,\bar{1}\,0]$ slip system. c2) In addition to the basal slip, an additional non-basal slip was activated. c3) Another slip system activated in the SmCo$_3$ phase. d1) Activation of the $(0\,0\,0\,1)[2\,\bar{1}\,\bar{1}\,0]$. d2-d3) Other identified slip systems in the Sm$_2$Co$_{17}$ phase.
}
\label{fig3}
\end{figure*}

\subsubsection{SmCo\texorpdfstring{\textsubscript{2}}{SmCo2}}
Spread over five grains of different orientations, 20 micropillars were compressed in this phase. Because of the limited grain size, only a small number of pillars per grain could be milled. Two orientations were chosen to enable compression tests close to the $\{1\,1\,1\}$ plane normal, while the rest were more strongly tilted from that plane. This approach aimed to activate non-$\{1\,1\,1\}$ slip in some orientations while ideally yielding slip on the $\{1\,1\,1\}$ plane in others. Yet, not all of the 20 pillars deformed in a manner suitable for slip system analysis. Some pillars cracked, crumbled, or showed too slight deformation to provide reliable insights into the slip plane or direction. Nonetheless, 16 out of 20  pillars could be assigned with a slip system. The pillars in orientations close to the $[0\,0\,1]$ corner of the standard triangle were correlated with slip on a $\{0\,1\,1\}$ plane, with slip directions $[\bar{1}\,1\,1]$ and $[0\,\bar{1}\,1]$ and Schmid factors of 0.49 and 0.43, respectively. In addition, the slip system $(1\,\bar{1}\,1)[\bar{1}\,0\,1]$ was observed in one instance with an orientation closest to $[0\,0\,1]$ corner (Figure \ref{fig3}a2). The pillar shown in  Figure \ref{fig3}a3 yielded a $(1\,1\,2)[2\,0\,\bar{1}]$ slip with $m=0.50$. For one orientation closest to the $[\bar{1}\,1\,1]$ corner we indexed a $\{1\,1\,n\}$ slip, with a slip system of $(1\,\bar{6}\,\bar{1})[\bar{1}\,0\,\bar{1}]$ ($m=0.49$). 

\subsubsection{SmCo\texorpdfstring{\textsubscript{5}}{SmCo5}}
Two orientations with different tilt angles from the basal plane were chosen, one expected to yield basal slip and the other expected to yield non-basal slip, based on the potential Schmid factors for basal slip in these orientations. In this phase, the identification of the slip systems is straightforward, as most pillars in each of the two orientations deformed consistently. Pillars in the basal orientation of the SmCo$_5$ phase exhibit deformation according to pyramidal II slip $(2\,\bar{1}\,\bar{1}\,\bar{1})[\bar{2}\,1\,1\,\bar{6}]$ with a Schmid factor of 0.44 (Figure \ref{fig3}b2). Most pillars in the orientation tilted from the basal orientation deformed through basal slip $(0\,0\,0\,1)[2\,\bar{1}\,\bar{1}\,0]$ with a Schmid factor of 0.45 (Figure \ref{fig3}b1), while only one instance showed pyramidal I slip $(0\,\bar{1}\,1\,1)[2\,\bar{1}\,\bar{1}\,0]$ with a comparable Schmid factor $m=0.45$ (Figure \ref{fig3}b3). More detailed investigations on the micropillar deformation in this phase have already been published in a previous study, where we also confirmed these findings via transmission electron microscopy (TEM) \cite{Stollenwerk2024}.

\subsubsection{SmCo\texorpdfstring{\textsubscript{3}}{SmCo3}}
For the SmCo$_3$ phase, 32 micropillars in 10 grains of different orientations were milled and compressed. Similar to the SmCo$_2$ phase, the grains are relatively small, measuring a few tens of microns across, with a needle-like crystal habit that allows the milling of 2--7 micropillars per grain at most. Depending on the orientation, different slip systems were activated, and as several orientations were tested, a variety of slip systems were identified. For the orientations 1, 3, 5, and 9 of the SmCo$_3$ phase, prismatic slip $\{0\,1\,\bar{1}\,0\}\langle2\,\bar{1}\,\bar{1}\,0\rangle$ was indexed (Figure \ref{fig3}c3) with Schmid factors of $m=0.42, 0.45, 0.44,$ and 0.34, respectively. Orientations in the $[0\,1\,\bar{1}\,0]$ corner of the triangle exhibited pyramidal slip along the $\{1\,0\,1\,\bar{1}\}$ planes. In two orientations $\langle a\rangle$ slip was observed, while in another orientation, deformation appeared to occur by $\langle c+a\rangle$ slip, with corresponding Schmid factors around $m= 0.40$. In the latter orientation, deformation was equally likely to follow the prismatic system $(0\,1\,\bar{1}\,0)[2\,\bar{1}\,\bar{1}\,0]$ with a Schmid factor comparable to the pyramidal system (i.e., $m=0.46$). Basal slip in this phase was less easily activated, with only two orientations with a tilt angle of about 50° from the $[0\,0\,0\,1]$ plane contained pillars that deformed in the $(0\,0\,0\,1)[2\,\bar{1}\,\bar{1}\,0]$ slip system.  In these cases, additional slip systems were present, but as basal slip was the most pronounced in the SE images (Figure \ref{fig3}c1 and c2), it was considered the dominant plastic event responsible for the yielding and most closely related to the corresponding flow stress.

\subsubsection{Sm\texorpdfstring{\textsubscript{2}}{2}Co\texorpdfstring{\textsubscript{17}}{17}}
Thirty pillars from three different grain orientations of Sm$_2$Co$_{17}$ were selected for micropillar compression tests. To perform nanomechanical testing on different crystallographic planes, a basal orientation and two inclined orientations with inclination angles of 63° and 67° were chosen. The basal orientation served as a reference for potentially activating non-basal slip. The two inclined orientations were rotated around the crystallographic $c$-axis at different angles to promote basal slip. The analysis of deformed micropillars shows very consistent results within each grain. In one of the tilted orientations, as shown in Figure \ref{fig3}d1, the compressed pillars exhibited basal slip on the $(0\,0\,0\,1)[2\,\bar{1}\,\bar{1}\,0]$ system, with a Schmid factor of 0.38. In the other tilted orientation, the slip occurred along a prismatic plane in the $(\bar{1}\,\bar{1}\,2\,0)[1\,\bar{1}\,0\,0]$ slip system, with a Schmid factor of 0.39 (Figure \ref{fig3}d2). For the basal orientation, deformation primarily occurred on pyramidal I slip systems $(\bar{1}\,1\,0\,1)[1\,\bar{1}\,0\,2]$ and $(\bar{1}\,0\,1\,1)[1\,0\,\bar{1}\,2]$, both with Schmid factors approximately 0.50 (Figure \ref{fig3}d3). Throughout the sample, a secondary phase was observed via SE imaging which was not detected in the EBSD mapping. Although we did not definitively identify this phase, it is likely the SmCo$_5$ phase, which is often observed alongside Sm$_2$Co$_{17}$ \cite{Buschow.1973,Campos2000} and exhibits similar Kikuchi patterns. This similarity could explain why the EBSD mapping missed this phase. Furthermore, these phases crystallize coherently, and the resulting stresses may cause lattice distortions that affect the Kikuchi patterns, making phase separation difficult without dedicated EBSD pattern simulation and analysis.

\subsection{Atomic-scale Modelling}

The material properties, including lattice constants and elastic properties, of the Sm-Co intermetallic phases were calculated using DFT and the EAM potential (see Table S1 in the SI). The EAM-calculated properties showed good agreement with the DFT results. Specifically, the deviations of the EAM-relaxed lattice constants were below 3.5\% compared to the DFT-relaxed values for the four Sm-Co phases. The deviations of the EAM-calculated elastic moduli were below 20\% for SmCo$_2$, SmCo$_3$, and SmCo$_5$, and ranged between 20\% and 30\% for Sm$_2$Co$_{17}$. Both the shear modulus and Young's modulus increased with higher Co content, as predicted by DFT and the EAM potential, while the Poisson's ratio decreased with increasing Co content.

The basal and $\{1\,1\,1\}$ slip systems in SmCo$_2$, SmCo$_3$, SmCo$_5$, and Sm$_2$Co$_{17}$ were theoretically investigated using the GSFE approach where the upper layers of the crystal were rigidly shifted along directions on the basal or $\{1\,1\,1\}$ plane to simulate crystallographic slip. Depending on the Sm-Co phase, different interlayers can facilitate basal slip, as shown in Figure \ref{fig1}. There are 2, 1, 3, and 3 unique interlayers along the basal planes of SmCo$_2$, SmCo$_5$, SmCo$_3$, and Sm$_2$Co$_{17}$, respectively. 

The $\gamma$-surfaces were generated for the unique basal interlayers of each Sm-Co crystal using the EAM potential, see Figures \ref{fig4} and S6. The $\gamma$-lines of energetically favorable pathways for the Sm-Co phases, highlighted via vectors on the $\gamma$-surfaces, were calculated using DFT (Figure \ref{fig4}). The GSFE profiles of the DFT-relaxed configurations were reevaluated using the EAM potential to assess the performance of this semi-empirical potential (Figure S7). The EAM potential showed good agreement with the DFT results. Since the DFT setup requires convergence tests on the z-dimension to minimize the interaction effects of the slip plane with non-periodic boundaries, which are computationally expensive for all Sm–Co systems, further comparisons of the energy barriers for basal slip among different Sm–Co phases were conducted using the EAM potential on atomistic configurations with a sufficiently large z-dimension (Figure S8), as detailed in Section \ref{modelmethods}.

The metastable states along the energetically favorable pathways were identified as stacking fault states, as shown in Figure S9. Despite the energy differences between the DFT and EAM-relaxed stacking fault configurations, they exhibit similar atomic arrangements at identical local minima. The following paragraphs introduce the energetically favorable pathways on $\gamma$-surfaces and the associated stacking fault states for each intermetallic phase.

\subsubsection{SmCo\texorpdfstring{\textsubscript{2}}{SmCo2}}

\begin{figure*}[hbpt!]
\centering
\includegraphics[width=0.9\textwidth]{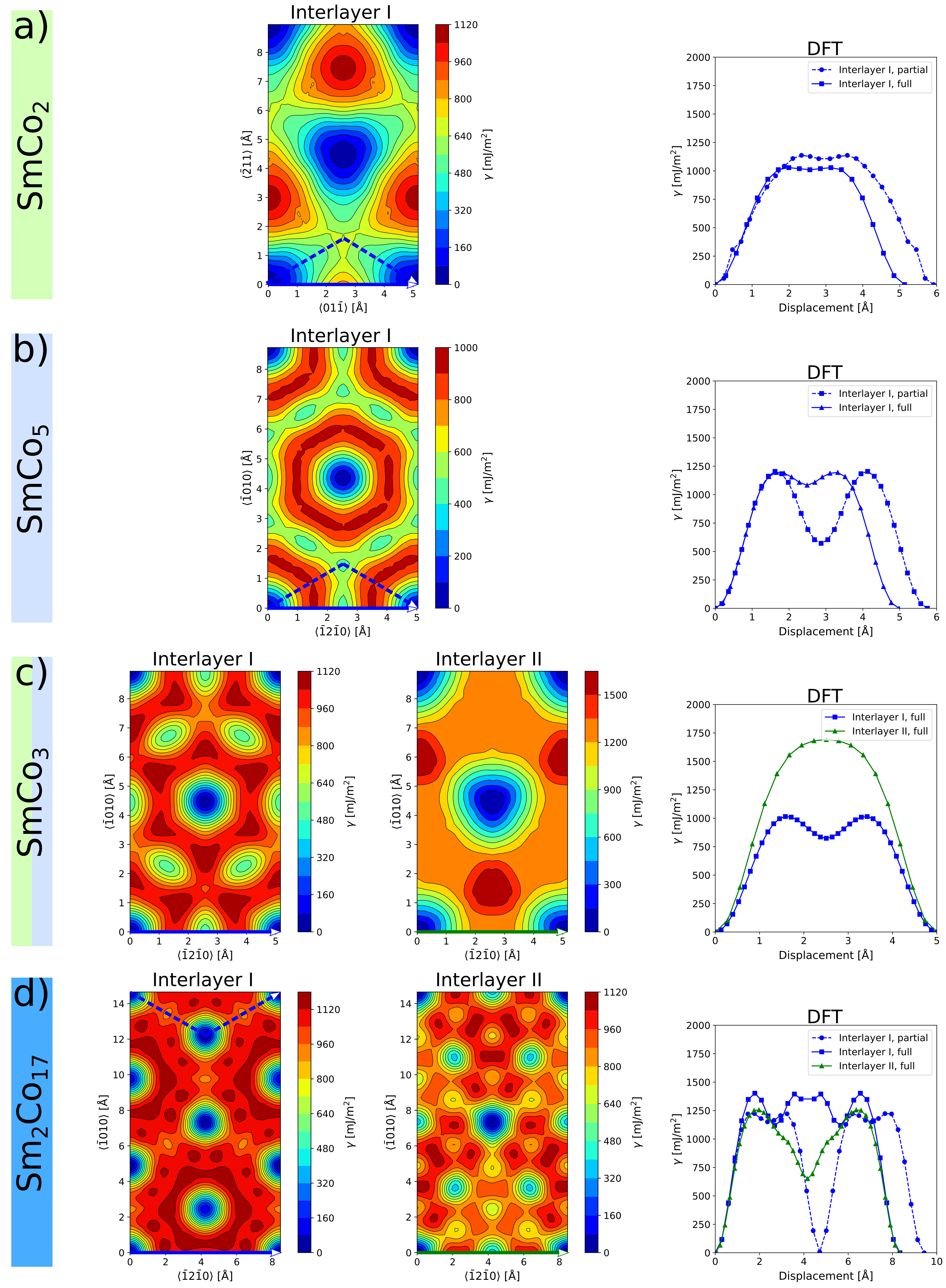}
\caption{Generalized stacking fault energy ($\gamma$) of the $\{1\,1\,1\}$ or basal plane in a) SmCo$_{2}$, b) SmCo$_{5}$, c) SmCo$_{3}$, and d) Sm$_{2}$Co$_{17}$. The $\gamma$ surfaces were calculated using the EAM potential. The $\gamma$ lines along the slip directions indicated on the ($\gamma$) surfaces were calculated using DFT.}
\label{fig4}
\end{figure*}

In the C15 SmCo$_2$ phase, $\{1\,1\,1\}$ slip can occur either between the Sm-Co-Sm triple-layer and the Co Kagomé layer (interlayer I) or within the triple-layer itself (interlayer II). Crystallographic slip is more favorable at interlayer I, which possesses a larger interplanar distance (see Figures \ref{fig4}a and S6a). In this interlayer, slip can proceed either along the full slip direction ($\frac{1}{2}\langle0\,1\,\bar{1}\rangle$) or along the dissociated partial slip pathway ($\frac{1}{6}\langle\bar{1}\,2\,\bar{1}\rangle$). DFT calculations indicate that the energy barrier for the partial slip pathway is approximately 108 mJ/m$^2$ higher than that for the full slip direction. Conversely, EAM results show the opposite trend, with the energy barrier for the full slip direction being 201 mJ/m$^2$ higher than that for the dissociated pathway. 

Metastable states were observed along both full and partial slip pathways in the DFT calculations, while only the partial slip pathway exhibited metastable states in the EAM potential (see the atomic configuration of the stacking fault state in Figure S9a). This qualitative difference in the energy profiles along the partial slip pathway implies a limitation of the EAM potential. The small energy differences between the metastable states and the adjacent energy barriers suggest the instability of the stacking fault states. This indicates the full slip along $\frac{1}{2}\langle0\,1\,\bar{1}\rangle$ at interlayer I is dominant in SmCo$_2$, as corroborated by both DFT and EAM approaches. The energy barriers for the basal slip events at interlayer I in SmCo$_2$ range from 648 to 826 mJ/m$^2$ along the EAM-relaxed GSFE profiles, representing the lowest basal slip barrier among the four Sm-Co intermetallics (see Figure S8 in the SI).

Due to the small interplanar distance between the Sm and Co layers within the triple-layer, crystallographic slip at interlayer II is highly unfavorable (Figure S6a). Instead, plasticity within the triple-layer is likely to occur through thermally activated synchro-shear slip, a high-temperature mechanism \cite{Xie2023,Xie2023a}. Therefore, basal plasticity at room temperature, corresponding to just over 20\% of the melting temperature of SmCo$_2$ \cite{Okamoto.2011}, is predominantly mediated by $\langle0\,1\,\bar{1}\rangle$ full slip at interlayer I, as supported by both DFT and EAM results, aligning with experimental observations.

\subsubsection{SmCo\texorpdfstring{\textsubscript{5}}{SmCo5}}

In the SmCo$_5$ structure, another fundamental building block of the Sm–Co intermetallics, there is only one basal interlayer. This interlayer has been previously investigated using the same EAM potential and DFT parameters by Stollenwerk et al. \cite{Stollenwerk2024}. Slip can occur along the full slip direction ($\frac{1}{3}\langle\bar{1}\,2\,\bar{1}\,0\rangle$) or through the dissociated $\frac{1}{3}\langle\bar{1}\,1\,0\,0\rangle$ directions, with similar energy barriers of 1195 and 1204 mJ/m$^2$ in DFT (1025 and 958 mJ/m$^2$ using the EAM potential), see Figure \ref{fig4}b and Figure S7b. Both DFT and EAM approaches predict metastable stacking fault states along the partial and full slip pathways (see the stacking fault configurations in Figure S9c-d). The basal slip barriers along the EAM-relaxed GSFE profiles range from 905 to 924 mJ/m$^2$, which are significantly higher than the energy barriers in the SmCo$_2$ phase (see Figure S8 in the SI).

\subsubsection{SmCo\texorpdfstring{\textsubscript{3}}{SmCo3}}

The SmCo$_3$ structure, comprising the fundamental building blocks of both the SmCo$_2$ and SmCo$_5$ crystals, has three unique basal interlayers labeled in Figure \ref{fig1}c. Although the SmCo$_2$ phase exhibited a lower energy barrier for basal slip compared to the SmCo$_5$ phase, the interlayer facilitating basal slip with the lowest energy barrier is the one in the SmCo$_5$ structure (labeled interlayer I) belonging to the $(0\,0\,0\,1)\langle\bar{1}\,2\,\bar{1}\,0\rangle$ slip system, based on DFT calculations and atomistic simulations (see Figure \ref{fig4}c). Along the $\langle\bar{1}\,2\,\bar{1}\,0\rangle$ slip pathway at interlayer I, a metastable stacking fault state was observed, with the stacking fault structure as illustrated in Figure S9b. The energy barrier for the same slip direction between the Sm-Co-Sm triple-layer and the Co Kagomé layer (labeled interlayer II) in the Laves crystal building block is significantly higher, even compared to the equivalent interlayer in the C15 SmCo$_2$ phase. 
The crystallographic slip between the triple-layer belonging to the SmCo$_2$ building block exhibited the highest energy barrier, making it the least favorable slip system similar to the SmCo$_2$ phase (Figure S6b). Therefore, for SmCo$_3$, the full slip at interlayer I is considered the dominant basal plastic event, with an energy barrier of 1031 mJ/m$^2$ obtained from the EAM-relaxed GSFE profile (Figure S8 in the SI). 

\subsubsection{Sm\texorpdfstring{\textsubscript{2}}{2}Co\texorpdfstring{\textsubscript{17}}{17}}

In Sm$_2$Co$_{17}$, three unique interlayers on the basal plane can be identified, as illustrated in Figure \ref{fig1}d. Besides interlayer III, which has the smallest interplanar distance, displacements along the full and partial slip pathways at interlayer I and II exhibited similar energy barriers (see Figures \ref{fig4}d and S6c). At interlayer I, the displacement along the partial slip pathway $\langle0\,1\,\bar{1}\,0\rangle$ exhibited a slightly lower energy barrier compared to the full slip direction $\langle\bar{1}\,2\,\bar{1}\,0\rangle$, according to both DFT and the EAM potential. A stable stacking fault with a stacking fault energy of only a few mJ/m$^2$ was identified after the partial slip, see the stacking fault structure in Figure S9f. The displacement along the full slip direction at interlayer II exhibited a similar energy barrier, where a metastable stacking fault was identified halfway along the slip pathway (see Figure S9e). For Sm$_2$Co$_{17}$, the energy barriers of basal slip events range from 1012 to 1110 mJ/m$^2$, obtained from the EAM-relaxed GSFE profiles, which is comparable to the SmCo$_3$ phase, with the upper bound slightly higher than that of SmCo$_3$ (see Figure S8 in the SI).

\section{Discussion}
\subsection{Mechanical Properties of Structurally Related Sm-Co Phases}

\begin{figure*}[hbt!]
\centering
\includegraphics[width=1\textwidth]{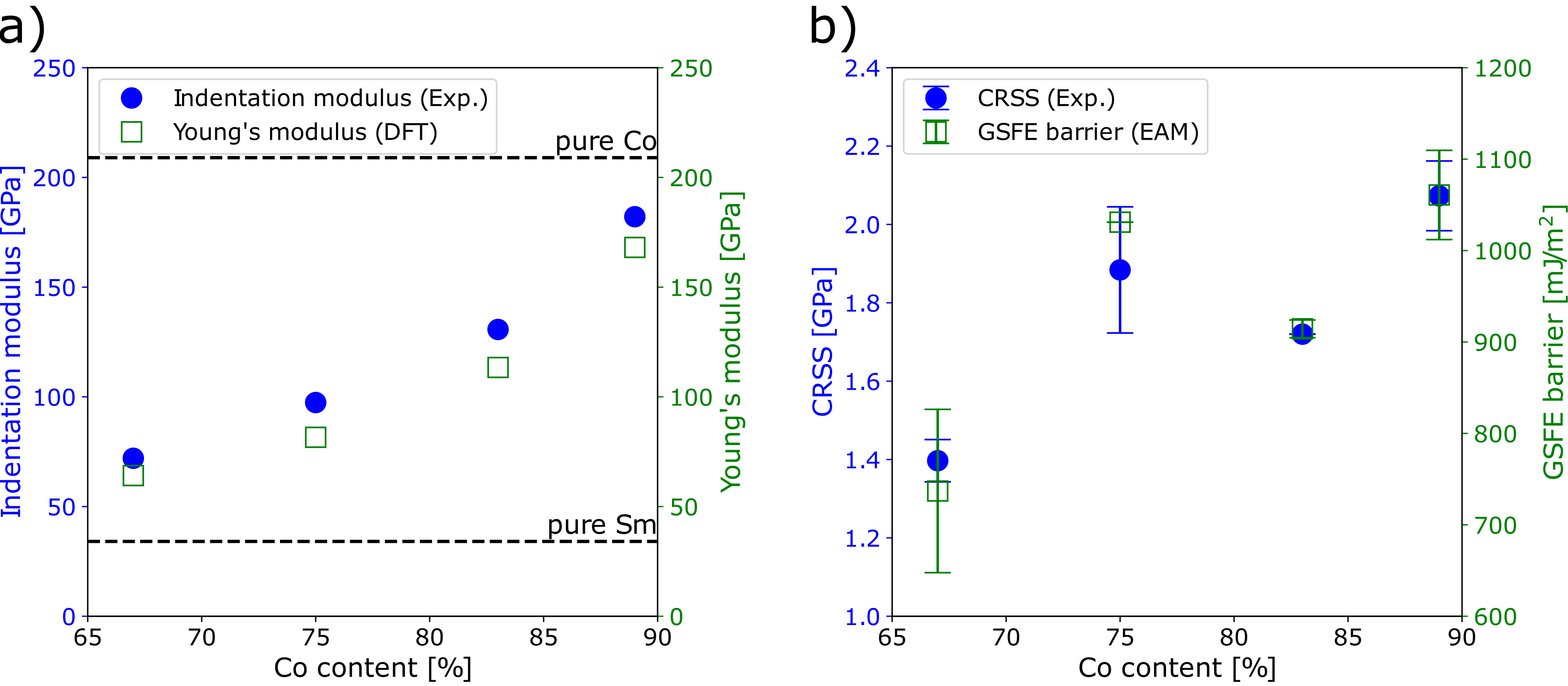}
\caption{Comparisons of a) experimentally measured indentation moduli and DFT-calculated Young's moduli of Sm-Co intermetallics, and b) critical resolved shear stresses (CRSS) obtained from micropillar compression experiments and generalized stacking fault energy (GSFE) barriers calculated using the EAM potential.}
\label{fig5}
\end{figure*}

The global mechanical properties, including hardness and indentation modulus, of the structurally related Sm-Co phases were measured using nanoindentation. The hardness values do not differ much from each other and range between 6 to 9 GPa across the four Sm-Co phases. Although there is a slight trend of increasing hardness with higher Co content, it is not pronounced. In contrast, the indentation modulus exhibits a more significant variation, which correlates with Co content. Specifically, the SmCo$_2$ phase has the lowest modulus at 72 ± 5 GPa, followed by SmCo$_3$ and SmCo$_5$ with indentation moduli of 98 ± 4 GPa and 131 ± 4 GPa, respectively. The Sm$_2$Co$_{17}$ phase displays the highest indentation modulus at 182 ± 3 GPa. This trend aligns well with the Young's modulus values calculated using DFT (see Figure \ref{fig5}a), which fall between the Young's moduli of Sm and Co, reported as 34.1 GPa and 209 GPa, respectively \cite{Samsonov.1968} (indicated by dashed horizontal lines in Figure \ref{fig5}a). 

Limited literature data is available for precise comparisons of hardness and modulus values for each of the four Sm-Co phases. The majority of available studies deal with the SmCo$_5$ phase. However, these studies often investigate the mechanical properties under specific circumstances or sample shapes, for example, investigations on thin films \cite{Kuru.2017}, sintered SmCo$_5$ magnets \cite{Singh2015} or compositionally altered samples \cite{Chikova2020} . Additionally, there are different ways to determine the mechanical properties, as acoustic methods are used to determine elastic moduli \cite{Doane.1977} and variations of indentation tests (Vickers and Knoop hardness) are used to determine hardness \cite{Singh2015,McCurrie.1971}. As a result, there is a wide range of different values for hardness and elastic modulus in SmCo$_5$. Hardness values range from 3.47 ± 0.07 GPa in thin films \cite{Kuru.2017} to 5.865 GPa for sintered magnets \cite{Singh2015,McCurrie.1971}. The elastic modulus values range from 43.09 ± 1.60 GPa \cite{Kuru.2017} up to 166.4 GPa \cite{Doane.1977}, depending on method, sample orientation and sample type. The values for reduced modulus of  131 ± 4 GPa we determined for the SmCo$_5$ phase are in good agreement with the literature and if it is considered that literature values for hardness are often determined from sintered magnets it is plausible that our hardness values (8.7 ± 0.2 GPa) lay a little higher than the literature values. 
Lu et al. \cite{Lu2009} presented an elastic modulus of 190.5 GPa for nanocrystalline Sm$_2$Co$_{17}$, consistent with our value of 182 ± 3 GPa, although their reported microhardness of 13.72 GPa is significantly higher than our measurements. This discrepancy might be attributed to a size effect at shallow depths of the material, as the tested nanocrystalline grains are claimed to be roughly 20 nm in size  \cite{Durst2008}. 
For pure SmCo$_2$ studies determine the Young's modulus to be around 60 GPa \cite{Klimker.1978}, while it is also suggested that a strong decline in elastic modulus occurs for temperatures below room temperature \cite{Klimker.1979}. 
SmCo$_3$ phases, no comparable studies on hardness and elastic modulus were found, which is why a study like the present one, that systematically determines the mechanical properties of the phases in the Sm-Co system is crucial for the comparability between each other and also other intermetallic systems.  Hence, the stated values derived from our measurements appear reasonable as they follow a consistent trend with increasing Co content, and our experimental values are supported by theoretical calculations (see Figure \ref{fig5}a).  Although the hardness and indentation modulus values follow this clear trend  (Figure \ref{fig5} and Table S2 in the SI), a sufficient deduction from mechanical properties of the building blocks to the complex phases cannot be made. If the mechanical properties were inherited from the building blocks we would expect that the values for SmCo$_3$ and  Sm$_2$Co$_{17}$ can be derived from the properties of SmCo$_2$ and SmCo$_5$. While this is true for the SmCo$_3$ phase, the values for  Sm$_2$Co$_{17}$ are much higher than the building block - SmCo$_5$ - it contains. This means that there must be other underlying factors that have an influence.
 
\subsection{Deformation Mechanisms in Fundamental Building Blocks}

Among the four investigated Sm-Co phases, SmCo$_2$ and SmCo$_5$ serve as fundamental building blocks for the structurally complex SmCo$_3$ and Sm$_2$Co$_{17}$ phases. 

In the SmCo$_2$ phase, slip trace analysis after nanoindentation revealed that the activation frequencies of indexed slip planes did not vary significantly across different orientations. The $\{1\,1\,2\}$ slip traces were most frequently observed, while slip traces for the $\{1\,1\,0\}$ and $\{1\,1\,1\}$ planes were less common, and $\{1\,0\,0\}$ slip traces appeared least frequently. From the micropillar compression tests, the $\{1\,1\,1\}\langle1\,\bar{1}\,0\rangle$ slip exhibited the lowest CRSS of 1.40 ± 0.05 GPa, followed by the $\{1\,1\,0\}\langle1\,\bar{1}\,0\rangle$ slip at 1.72 ± 0.10 GPa and the $\{1\,1\,2\}\langle1\,\bar{1}\,0\rangle$ slip at 1.99 ± 0.10 GPa. Additionally, in one micropillar compression orientation, a higher indexed slip system, $(1\,\bar{6}\,1)\langle1\,\bar{0}\,1\rangle$, with a CRSS of 2.06 ± 0.13 GPa, was identified. Given that there are more conjugated $\{1\,1\,2\}$ planes (12 in total) than conjugated $\{1\,1\,1\}$ planes (four in total), the higher activation frequency of $\{1\,1\,2\}$ slip in indentation is reasonable despite its higher activation barrier. The similar CRSS values of $\{1\,1\,2\}$ and $\{1\,1\,6\}$ slip suggest the possibility of $\{1\,1\,n\}$ slip in this phase, as proposed recently by Freund et al. in their investigation of an isostructural Laves phase \cite{Freund2023, Freund2024}.

The deformation mechanisms of Laves phases, particularly the synchro-shear slip mechanism on $\{1\,1\,1\}$ and basal planes, have been extensively studied through both experiments \cite{Chu1994,Chisholm2005,Kazantzis2007} and atomic-scale modeling \cite{vedmedenko2008,guenole2019}. More recently, kink-pair nucleation and propagation have been demonstrated as mechanisms for synchro-shear dislocation motion using atomistic simulations \cite{Xie2023a}. Additionally, the thermally activated nature of synchro-shear dislocation motion has been investigated, highlighting that thermal fluctuations are indispensable in activating synchro-shear slip \cite{Xie2023}. Consequently, synchro-shear slip is often reported as the dominant deformation mechanism in high-temperature processes but cannot be operated to mediate plasticity at room temperature, which aligns with our experimental observation of the lower activation frequencies of $\{1\,1\,1\}$ compared to other $\{1\,1\,n\}$ planes. Our observation of the activation of $\{1\,1\,n\}\langle1\,\bar{1}\,0\rangle$ slip systems also correlates well with a recent nanoindentation and TEM study on a C15 Ca-Al-Mg Laves phase suggesting that deformation at room temperature is not limited to the $\{1\,1\,1\}$ plane but also occurs on higher indexed $\{1\,1\,n\}$ planes via full dislocation slip \cite{Freund2024}. It has also been demonstrated via atomistic simulations that $\{1\,1\,1\}$ slip is likely to occur at the interlayer between the triple-layer and Kagomé layer along the $\langle1\,\bar{1}\,0\rangle$ direction at room temperature, with an energy barrier comparable to that of other $\{1\,1\,n\}$ slip systems \cite{Freund2024}. Our study of deformation mechanisms of the SmCo$_2$ fundamental building block are therefore in good agreement with the mechanisms identified on other C15 Laves phases, revealing a strong transferability of insights between isostructural phases containing different element combinations in this size driven topologically close packed phase.

The indentation results for the SmCo$_5$ phase show that the most prominent slip system is pyramidal slip, while only a small portion of the overall slip traces are basal slip traces. This observation is primarily affected by the anisotropic generation of traces throughout the chosen orientation, with mainly one orientation, which was specifically chosen to yield basal slip based on the Schmid factor, contributing to the total number of observed basal slip traces. By likewise carefully selecting the orientation for the micropillar compression tests, we were able to trigger pyramidal II slip as well as basal slip along the $(2\,\bar{1}\,\bar{1}\,\bar{1})\langle2\,\bar{1}\,1\,\bar{6}\rangle$ direction with a CRSS of 1.64 ± 0.4 GPa and the $(0\,0\,0\,1)\langle2\,\bar{1}\,\bar{1}\,0\rangle$ direction with a CRSS of 1.72 GPa, respectively \cite{Stollenwerk2024}.
Given that there are six conjugated pyramidal II planes, one would expect six times more pyramidal II slip traces than basal slip traces, considering the similar CRSS levels. In the SmCo$_5$ phase, basal slip traces are not as frequent as one would expect from a hexagonal crystal. Studies elsewhere on the SmCo$_5$ phases have reached controversial conclusions. While some suggest that deformation occurs through amorphous shear band formation \cite{Luo2020}, others conclude that dislocation movement occurs along distinct crystallographic planes \cite{Fidler1978}, which is consistent with our previous study \cite{Stollenwerk2024}. Furthermore, studies on a LaNi$_5$ phases with this structure type suggest, that active slip systems in this building block phase are basal $(0\,0\,0\,1)\langle1\,1\,\bar{2}\,0\rangle$  and prismatic $(1\,\bar{1}\,0\,0)\langle1\,1\,\bar{2}\,0\rangle$, which are slip systems that are in good correspondence with our findings \cite{Inui.1998}. In general, our data agrees with expectations drawn from the information available in literature on isostructural phases.

\subsection{Deformation Mechanisms beyond Fundamental Building Blocks}

\begin{figure*}[hbt!]
\centering
\includegraphics[width=1\textwidth]{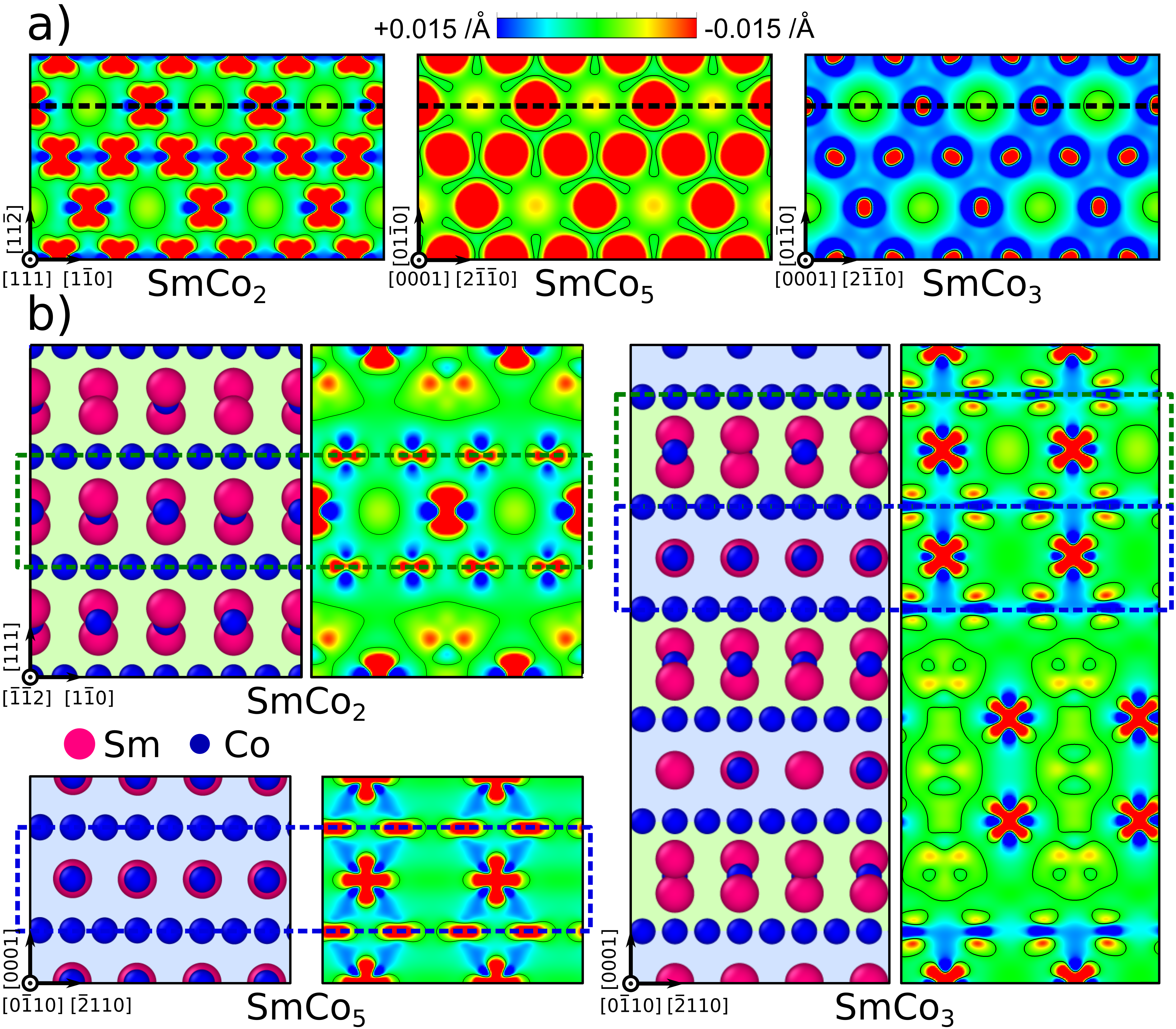}
\caption{Charge density difference profiles of a) Co Kagomé-nets along $\{1\,1\,1\}$ or basal planes and b) $\{1\,1\,2\}$ or prismatic atomic layers (annotated as dotted lines in a)) in the SmCo$_2$, SmCo$_5$, and SmCo$_3$ crystals. Regions of charge concentration and deficiency are displayed in red and blue respectively, and green areas represent a metallic or delocalized bonding environment.  The fundamental building blocks in the SmCo$_3$ crystal are indicated by different background colors and the equivalent $\{1\,1\,2\}$ and prismatic atomic layers are marked using dashed boxes.}
\label{fig6}
\end{figure*}

Having considered the active deformation mechanisms in the two fundamental building block phases, we turn to the structurally more complex crystals. In particular, we discuss how the observed mechanisms are affected by the stacking of the building blocks into larger unit cells, such as in the SmCo$_3$ phase, which is composed of the SmCo$_2$ and SmCo$_5$ building blocks stacked along the $\langle1\,1\,1\rangle$ and $\langle0\,0\,0\,1\rangle$ directions, respectively. From a purely geometric point of view, it is reasonable to expect that the mechanical responses and deformation mechanisms of these structurally complex crystals, particularly in directions parallel to the joint Kagomé atomic layers between fundamental building blocks, are closely correlated with the properties of the individual building blocks. An example can be seen in the topologically close-packed $\mu$ phase, which consists of C15 Laves phase and Zr$_4$Al$_3$ building blocks stacked along the joint Kagomé atomic layers, where the elastic properties and basal plasticity can be interpreted and even predicted based on the characteristics of the fundamental building blocks \cite{Schroders2018,Schroders2019,Luo2023,luo2023plasticity}. The given directionality of the stacking along one axis may also inhibit those mechanisms crossing the stacking plane. In case of the $\mu$-phase, this has in fact been shown in the form of a strong preference for basal slip \cite{luo2023plasticity, Luo2023}, compared with the activation of many slip planes in the Laves phases \cite{Freund2023, Freund2024}. In addition, changes in bonding or lattice spacing may occur as the building blocks are combined, which would again be expected to have a strong effect but which are less easily predicted from prototypes of unit cells.

Our study on the structurally related Sm-Co phases reveals that a more complex scenario indeed occurs, where mechanisms and critical stresses cannot be deduced directly from geometric considerations alone. While the elastic properties of the SmCo$_3$ phase do follow a rule of mixtures when considering the SmCo$_2$ and SmCo$_5$ phases, the plastic deformation mechanisms do not align with a simple interpretation of identical building block plasticity and a constraint in non-stacking direction. Despite SmCo$_2$ being much softer along the $\{1\,1\,n\}$ plane (CRSS: 1.4 GPa) than SmCo$_5$ along the basal plane (CRSS: 1.7 GPa), as demonstrated here by micropillar compression as well as GSFE calculations, the SmCo$_3$ phase deforms via mechanisms associated with the SmCo$_5$ building block, but with an even higher CRSS than SmCo$_5$ (1.88 GPa). This discrepancy cannot be explained by geometric factors alone, as the interplanar distance over the Burgers vector ($d/b$) ratios of equivalent slip systems remain similar in the SmCo$_2$ and SmCo$_5$ phases and their corresponding building blocks in SmCo$_3$. Specifically, the $d/b$ ratios for full slip along the triple-layer and Kagomé layer in SmCo$_2$ and SmCo$_3$ are 0.307 and 0.293, respectively, too small to give a significant difference in lattice resistance beyond 1--2\% in spite of entering an exponential function in estimates of the Peierls stress -- yet the mechanical response we find here differs significantly. 

This initially unexpected behavior arises because the local bonding environment of the slip plane is significantly altered when these building blocks are stacked together. The change in the local bonding environment can be observed by comparing the charge density difference profiles of the SmCo$_2$ and SmCo$_5$ building blocks with that of SmCo$_3$ (see Figure \ref{fig6}). The charge density difference profiles of the Co Kagomé layers in SmCo$_2$, SmCo$_5$, and SmCo$_3$, as shown in Figure \ref{fig6}a, indicate that the level of electron localizability varies among these Sm-Co phases. The Kagomé layer in SmCo$_5$ is characterized by predominantly metallic bonding, as indicated by the circular charge concentration areas, which lack directional preference. In contrast, the Kagomé layers in SmCo$_2$ and SmCo$_3$ exhibit regions of charge deficiency, implying more localized electrons and a correspondingly more directional bonding nature. This variation in electron distribution highlights how the local bonding environment, influenced by adjacent atomic layers, alters the electronic properties of the structurally identical Co Kagomé layers across different Sm-Co phases. 

Viewing the charge density difference profiles within a slice of the $\{1\,1\,2\}$ or prismatic planes of SmCo$_2$, SmCo$_5$ and SmCo$_3$ in Figure \ref{fig6}b provides a clearer picture of the interaction at the $\{1\,1\,1\}$ or basal interlayers. Based on our calculations, electrons surrounding the Sm atoms are delocalized in all three crystals, blending into the background charge within the crystal. Although the limitation of the Sm PAW potential used in this work may risk underestimating the Sm 4$f$ to Co 3$d$ exchange interaction and the localization of the $f$-orbital states \cite{graanas2012charge,jeffries2014robust,das2019anisotropy}, our charge density profile is consistent with other studies where the Sm 4$f$ orbitals are explicitly described \cite{das2019anisotropy,mao2019structural}. In all three phases, the Co atoms between the Kagomé layers show directional bonding with the Co atoms within the Kagomé layers, which is attributed to covalent-like Co-Co bonding \cite{mao2019structural}. 

The low energy barrier for the crystallographic slip along $\{1\,1\,1\}\langle1\,\bar{1}\,0\rangle$ in SmCo$_2$ can be attributed to the parallel alignment between the charge-concentrated/bonding regions and the slip plane, indicating relatively low hybridization of the bonding states between the Co atoms in the Kagomé layer and those in the triple-layer. In contrast, in SmCo$_3$, the modulation of the Kagomé states works in tandem with the localization of Co electrons within its adjacent building blocks, leading to increased overlap or hybridization between neighboring atomic layers perpendicular to the basal slip direction. 
As a result, basal slip in the SmCo$_2$ building block of the SmCo$_3$ phase becomes significantly more difficult than in the pure SmCo$_2$ phase, to the extent that slip within the SmCo$_5$ building blocks becomes the prevalent basal plasticity mechanism in SmCo$_3$. These findings highlight the limitations of using fundamental building blocks alone to predict the properties of structurally complex crystals. The interactions between the building blocks and the resulting changes in bond nature must be considered to understand and accurately predict the deformation behavior of these structurally related materials.

With respect to the balance of basal versus non-basal slip, SmCo$_3$ exhibits nearly half of the activated slip traces as basal slip (48\%). Similarly, the Sm$_2$Co$_{17}$ structure, which is derived from the SmCo$_5$ structure by replacing one-third of the Sm atoms with a dumbbell of two Co atoms, exhibits not only a higher basal CRSS (20\%) compared to SmCo$_5$ due to more compact interplanar distances but also a significantly higher fraction of basal slip traces (25\%). Given the strong imbalance between available equivalent planes between the single basal and multiple non-basal planes for each candidate system, this result indicates that the non-basal plasticity in this crystal is largely suppressed. In contrast, the fundamental building blocks, SmCo$_2$ and SmCo$_5$, exhibit plasticity dominated by non-$\{1\,1\,1\}$ and non-basal slip systems (see Figure \ref{fig2}). The stacking sequence of these building blocks in SmCo$_3$ truncates the non-basal slip systems, thereby hindering their activation. As indicated above, similar phenomena have been reported in the $\mu$ phase at room temperature, where basal slip dominates deformation, while most non-basal slip mechanisms disrupt the Frank-Kasper packing of topologically close-packed phases \cite{luo2024non}. In contrast, most deformation in the C15 Laves phase, as one of the building blocks of the $\mu$ phase, is carried by non-$\{1\,1\,1\}$ dislocations \cite{Freund2024}. Another example can be found in the quasiternary system of Ni$_3$V–Co$_3$V–Fe$_3$V, where the stacking sequence of the building blocks can be controlled by the number of valence electrons per atom, which in turn affects the accessible slip systems and the material's ductility \cite{liu1984ductile}. 

\section{Conclusions}

In this study, we investigated the deformation behavior of structurally complex Sm-Co intermetallic phases, focusing on the relationships between fundamental building blocks -- SmCo$_2$ and SmCo$_5$ -- and their structurally related derivatives, SmCo$_3$ and Sm$_2$Co$_{17}$, using a combination of nanoindentation, micropillar compression, and atomic-scale modeling. Our findings reveal that the plastic deformation mechanisms in these structurally complex phases are not simply inherited from the building blocks. Instead, the local bonding environment plays a crucial role in altering the deformation behavior. Specifically, the interactions between adjacent atomic layers, particularly within Co Kagomé-nets, lead to changes in the bonding characteristics, which in turn influence the ease of crystallographic slip processes. 

Building on this and other studies, we therefore propose that stacking of smaller fundamental building blocks in more complex phases leads to different effects depending on whether the deformation occurs parallel to or across the stacking planes. Across the stacking plane, deformation mechanisms observed in the smaller cells of the building block phases tend to be suppressed, rendering slip across the stacking plane unfavourable. Within the stacking plane, the changes are more nuanced and involve alterations in critical stress as well as the balance of competing mechanisms.

We suggest that in order to predict how a stacked phase deforms parallel to the stacking plane, two key aspects should be considered: (1) changes in lattice and slip plane spacing relative to the fundamental building blocks, and (2) changes in local bonding environments due to varying charge density distributions within the stacking planes. Whether these aspects may be related to even simpler factors, such as atomic size ratios and electronegativity of the atoms within and adjacent to the stacking planes, remains to be explored.

Ultimately, we aim to accelerate the exploration and optimization of plastic properties across the vast intermetallic phase and composition space. With this work, we contribute to this goal by uncovering the diverse deformation mechanisms of fundamental building block phases, which serve as the foundation for many larger structures. Additionally, we provide initial guidelines for predicting how the properties of these building blocks can inform the behavior of larger unit cells, focusing on the simplest possible parameters, such as changes in compositional, structural, and electronic environments.

\section{Methods}\label{methods}
\subsection{Sample Preparation}
The Sm-Co samples with a total mass of 1 g each were synthesized from pure Co (Puratronic, purity 99.995\%) and Samarium (chempur, purity 99.99\%) in stoichiometric ratio. The elements were melted in an arc melter (Edmund Bühler MAM-1), turned, and re-melted at least three times to achieve better homogeneity. As samarium exhibits a significantly high vapor pressure \cite{Herrick1964}, an effort was made to not expose the material to the arc for too long so that the mass loss was kept to a minimum. Afterward, they were ground and polished utilizing an OP-S colloidal suspension for the final polish. The SmCo$_2$ and SmCo$_3$ samples did not yield any usable electron backscatter diffraction (EBSD) pattern with mechanical polishing, which is why they were additionally ion polished on a SM-09010 Cross Section Polisher (JEOL Ltd. (JP)) under argon atmosphere with 6 kV acceleration voltage and a current of 0.24 mA. As the ion beam hits the sample surface at a grazing angle from the side, a curtaining effect occurs that produces an undulating surface which can be inconsistently pronounced. During EBSD measurements it was apparent that the hexagonal and the rhombohedral polymorphs of Sm$_2$Co$_{17}$ are not easily distinguished.  As the samples were synthesised using an arc-melting furnace temperature could not be controlled to a level where a specific polymorph could be targeted. In this study, we therefore assume that through rapid cooling we retain the hexagonal polymorph, which also seamed more reasonable in the EBSD measurements. The SmCo$_5$ samples were synthesised and prepared during a previous study, which also reports on the micropillar compression tests and the corresponding analysis \cite{Stollenwerk2024}. 

\subsection{Electron Backscatter Diffraction}
For electron backscatter diffraction, a Focused Ion Beam (FIB)-SEM Helios Nanolab 600i (FEI (NL)) was used. EBSD patterns were collected through a Hikari camera (EDAX Inc.) at an acceleration voltage of 20 kV and a beam current of 2.7 nA. This way, an orientation map for each sample was determined in order to select orientations for nanomechanical tests.

\subsection{Nanoindentation}
For nanoindentation tests, we utilized an iNano indentation device from Nanomechanics Inc. (USA) equipped with a diamond Berkovich tip from Synton-MDP (CH). Tests were performed with a constant strain rate of 0.2 s$^{-1}$ until a maximum depth of 1000 nm or a maximum load of 45 mN was reached. The data was analyzed using the Oliver-Pharr method \cite{Oliver1992}. Hardness and indentation modulus values were determined for each of the four Sm-Co phases. The Poisson’s ratio and the elastic modulus were set to 0.07 and 1140 GPa for the diamond Berkovich tip. For the indentation analysis, Poisson’s ratios given in Table S1 in the Supplementary Information (SI) were used for the samples. 

To analyze the slip traces, the indents of the nanoindentation tests were imaged using secondary electron (SE) images (taken with a (FIB)-SEM Helios Nanolab 600i (FEI)). The traces around the indent were manually marked and compared to the angular orientation of traces that selected lattice planes would form when intercepting the sample surface using the MATLAB\textregistered{} code published by Gibson et al. \cite{Gibson2021,Pei2021}. To identify the most likely slip plane, a threshold angle of  3° was set, and the deviation angle between the marked trace and the calculated traces was recorded. Slip traces formed around an indent yield two-dimensional information of slip planes intercepting the sample surface. However, the actual tilt angle of the plane into the sample cannot be determined from this information alone. For instance, prismatic and pyramidal slip traces may appear similar in basal orientation. Consequently, this slip trace analysis only provides an estimation of the actual slip activation frequency.
Indentation results in the SmCo$_2$ phase had to be corrected for the samples surface tilt as the ion polishing induced a significant tilt of about 5° from the horizontal, causing the indenter tip to not penetrated the sample exactly perpendicular and distorting the indentation marks to non-equilateral triangles. We corrected the mechanical properties for this tilt according to the works of Kashani and Madhavan \cite{Kashani2011}, who performed finite element simulations showing that indentation with a Berkovich indenter tip on a tilted sample surface results in a 12\% error in hardness for a 5° tilt.

\subsection{Micropillar Compression}
Micropillars were cut in successive milling steps with different ion beam currents using FIB of a (FIB)-SEM Helios Nanolab 600i (FEI). A height-to-diameter ratio of 2:1 with a pillar height of 4 $\mu$m and a diameter of 2 $\mu$m was kept. Compression tests were conducted on a FT-NMT04 from FemtoTools (CH) equipped with a 200 mN flat punch sensor built into a TESCAN CLARA SEM (CZ). Micropillars in the SmCo$_5$ were compressed on an InSEM equipped with an InForce50 actuator (Nanomechanics Inc. (USA)) and a 5 $\mu$m flat punch from Synton-MDP (CH). In total 112 pillars were milled and compressed for the four different Sm-Co phases combined. In each phase, grains of different orientations were picked to enable the activation of slip along different crystallographic planes deliberately aiming in particular at the basal plane for the hexagonal/rhombohedral phases and the $\{1\,1\,1\}$ plane for the cubic Laves phase.

Using the local crystal orientation from the EBSD measurements, model visualizations of the slip geometries of the micropillars were calculated \cite{Pei2021}. After compression, the deformed pillars were imaged with an electron microscope at four rotation angles every 90° about the compression axis to get the best view of the slip plane and additionally from a top-down perspective to identify the slip direction. Afterward, the SE images of the pillar were compared to the precalculated compression models and visualizations of the slip plane and direction in VESTA (Visualization for Electronic Structural Analysis) \cite{Momma2011}. Furthermore, the Schmid factors for potential slip systems were calculated and taken into consideration when choosing the most likely slip systems.

\subsection{Atomic-scale Modelling}\label{modelmethods}
All atomic configurations in this study were constructed using Atomsk \cite{hirel2015} and visualized using the Open Visualization Tool (OVITO) \cite{Stukowski2010}.
GSFE lines and surfaces were computed by molecular statics simulations using the LAMMPS software \cite{Thompson2022}. These calculations involved incrementally shifting one-half of the crystal along specific slip directions on the corresponding slip planes. The Sm-Co embedded atom method (EAM) potential, developed by Luo et al. \cite{Luo2020,Luo2019}, was used to model the interatomic interactions. Periodic boundary conditions were applied in the x and y directions parallel to the slip plane, while the upper and lower atomic layers were fixed with a distance of 14 Å in the z-direction. The simulation structures maintained an aspect ratio of approximately 10, based on the z-direction length relative to the larger of the x- or y-lengths. The FIRE algorithm with a force tolerance of $10^{-8}$ eV/Å was used for relaxation in the direction perpendicular to the slip plane after each incremental shift \cite{Guenole2020}.

DFT calculations were conducted using the Vienna ab initio Software Package (VASP) \cite{Kresse1993,Kresse1994a}. The Kohn-Sham one-orbital wavefunctions were represented by planewave basis sets through the Projector Augmented Wave (PAW) potential \cite{Kresse1996a,Kresse1996b,Kresse1999}. Core electrons were frozen, and only valence electrons were involved in bonding, specifically Sm 5$s^2$, 6$s^2$, 5$p^6$, 5$d^1$ and Co 3$d^8$ 4$s^1$ (POTCAR versions November 2014 and August 2007, respectively). The kinetic cut-off energy for the planewave basis was set to 550 eV. The Perdew, Burke, and Ernzerhof (PBE) exchange-correlation functional based on the Generalized Gradient Approximation (GGA) was used \cite{perdew1996generalized}. Smearing at the Fermi level was applied using the first-order Methfessel-Paxton method \cite{methfessel1989high} with a value of 0.01. Electronic and geometric convergence thresholds were set at 10$^{-6}$ eV and 0.01 eV/Å, respectively. All calculations were spin-polarized. These calculation parameters are consistent with those used in our previous work \cite{Stollenwerk2024}.

The lattice parameters of SmCo$_2$, SmCo$_3$, SmCo$_5$, and Sm$_2$Co$_{17}$ were optimized using DFT. The full cell relaxation scheme for crystals with two independent lattice parameters involved an initial two-dimensional scan across the lattice parameters, during which only atomic positions were relaxed. From the resulting contour plot, a starting point for the energy-minimum lattice parameters was determined and further optimized using a full cell relaxation scheme (including optimization of lattice parameters, angles, and atomic positions). The Monkhorst-Pack $k$-point scheme was used to sample the Brillouin zone \cite{monkhorst1976special}. Gamma-centered $k$-point meshes of 12$\times$12 $\times$12, 8$\times$8$\times$1, 20$\times$20$\times$20, and 5$\times$5$\times$5 were used to optimize the bulk lattice constants of SmCo$_2$, SmCo$_3$, SmCo$_5$, and Sm$_2$Co$_{17}$, respectively. The lattice parameters of the Sm-Co crystals obtained using DFT fell within the range of the experimentally measured lattice constants \cite{Kanematsu1989,Harris1965,Buschow1968,nassau1960intermetallic,Buschow1968,Bertaut1965,Lihl1969,Song2009,Buschow1968}.

The magnetic configurations of the Sm-Co phases were systematically explored, as presented in Figure S1. Our investigation indicates that the lowest energy structures of all Sm-Co crystals studied correspond to a state in which the Co spin moments are antiparallel to the Sm spin moments, with total magnetic moments of 1.98, 3.62, 7.37, 26.5 $\mu_B$/f.u. for SmCo$_2$, SmCo$_3$, SmCo$_5$, and Sm$_2$Co$_{17}$, respectively. These results are consistent with current theories of magnetism in Sm-Co crystals \cite{kumar1988retm5,graanas2012charge,jeffries2014robust,soderlind2017prediction}. This work does not include the explicit description of Sm $f$ orbitals as valence electrons, as the GGA formulation does not adequately treat the localization of $f$-orbitals \cite{das2019anisotropy,graanas2012charge}. Spin-orbit coupling is also excluded, as it has little influence on the equilibrium volume \cite{jeffries2014robust}. Although our calculations likely over-delocalize Sm states and underestimate the Sm 4$f$ to Co 3$d$ coupling, this effect is not expected to qualitatively affect the Co-Co bonding characteristics discussed in this work.

The elastic tensors were calculated in DFT using the energy-strain approach. Distortion modes implemented by AELAS \cite{zhang2017aelas} were applied to the equilibrium structure, and an energy-strain fit was done to determine the quadratic coefficients and, thus the elastic tensor components. A minimum of 1000 $k$-points per reciprocal atom was used for all structures. The resulting list of the elastic constants is provided in Table S1 in the SI.

The atomic configurations and their dimensions for GSFE calculations of each Sm-Co structure using DFT are displayed in Figure S2. To ensure that atoms do not interact with its periodic image perpendicular to the slip plane, a vacuum layer of 13 Å was inserted in between. Benchmarking tests of the $k$-point mesh were performed and detailed in the SI (Figure S3). The DFT models consist of 1$\times$1$\times$3, 1$\times$1$\times$2, and 1$\times$1$\times$4 basal or $\{1\,1\,1\}$-oriented SmCo$_2$, SmCo$_3$, and Sm$_2$Co$_{17}$ supercells, respectively. Gamma-centered $k$-point meshes (using the Monkhorst-Pack scheme) of 3$\times$6$\times$1, 6$\times$3$\times$1, and 3$\times$2$\times$1 for the SmCo$_2$, SmCo$_3$, and Sm$_2$Co$_{17}$ structures, respectively, were employed to ensure that the $k$-points per reciprocal atom were at least 1000. All corresponding parameters for the GSFE calculation of SmCo$_5$ can be found in our previous work, where the set-up approach was identical \cite{Stollenwerk2024}.

The local bonding environments were characterized by charge density differences which indicate where charge is concentrated or deficient within the crystal. This is represented by the equation:
\begin{equation}
    \Delta \rho_\mathrm{Sm_nCo_m}  = \rho_\mathrm{Sm_nCo_m} - n\rho_\mathrm{Sm} - m \rho_\mathrm{Co},
\end{equation}
where $\rho_\mathrm{Sm_nCo_m}$ is the charge density of the bulk Sm-Co intermetallic phases, and $\rho_\mathrm{Sm}$ and $\rho_\mathrm{Co}$ are isolated single-atom charge densities of Sm and Co, respectively, within the same unit cells as the intermetallic phases. The charge density profiles were visualized using VESTA.

\section*{Acknowledgements}

T.S., N.Z.Z.U, Z.X., and S.K.K. are grateful for funding from the European Research Council (ERC) under the European Union’s Horizon 2020 research and innovation program (grant agreement No. 852096 FunBlocks). Additionally, P.C.H., Z.X., and S.K.K. acknowledge financial support from the German Research Foundation (DFG) through the SFB1394 Structural and Chemical Atomic Complexity – From Defect Phase Diagrams to Material Properties, project ID 409476157. The authors gratefully acknowledge the computing time provided to them at the NHR Center NHR4CES at RWTH Aachen University (project numbers p0020431 and p0020267). This is funded by the Federal Ministry of Education and Research, and the state governments participating on the basis of the resolutions of the GWK for national high performance computing at universities (www.nhr-verein.de/unsere-partner).

\section*{Data availability}
The datasets used and/or analysed during the current study available from the corresponding author on reasonable request.

\section*{Conflict of Interest}
The authors declare no conflict of interest.

\bibliographystyle{MSP}
\bibliography{main}

\clearpage

\onecolumngrid
\vspace*{8cm} 
\begin{center}
    {\Large \textbf{Supplementary Information}} \\
    \vspace{0.5cm} 
    {\Large Beyond Fundamental Building Blocks: Plasticity in Structurally Complex Crystals} \\
    \vspace{0.5cm} 
    {\Large Stollenwerk, et al.}
\end{center}

\vspace{1cm} 
\clearpage
\setcounter{figure}{0}
\setcounter{table}{0}
\renewcommand{\figurename}{Figure S}
\renewcommand{\tablename}{Table S}

\begin{table*}[!hpt]
\centering
\caption[]{\label{tab1}Properties of Sm-Co phases determined through DFT calculations and atomistic simulations using the EAM potential. $a\textsubscript{0}$ and $c\textsubscript{0}$: lattice parameters; $C\textsubscript{ij}$: elastic constants; $B$: bulk modulus (Hill approximation); $G$: shear modulus (Hill approximation); $E$: Young's modulus (Hill approximation); $\nu$: Poisson's ratio (Hill approximation).}
\centering
\scriptsize
\begin{tabular}{p{0.1\textwidth}p{0.05\textwidth}p{0.15\textwidth}p{0.05\textwidth}p{0.15\textwidth}p{0.05\textwidth}p{0.15\textwidth}p{0.05\textwidth}p{0.15\textwidth}}
\hline\hline
\addlinespace[0.1cm]
\multicolumn{1}{l}{} & \multicolumn{2}{c}{SmCo$_2$} & \multicolumn{2}{c}{SmCo$_5$} & \multicolumn{2}{c}{SmCo$_3$} & \multicolumn{2}{c}{Sm$_2$Co$_{17}$}\\
\cmidrule(lr){2-3}
\cmidrule(lr){4-5}
\cmidrule(lr){6-7}
\cmidrule(lr){8-9}
Properties & DFT & EAM & DFT & EAM & DFT & EAM & DFT & EAM\\
\addlinespace[0.1cm]
\hline
\addlinespace[0.1cm]
$a{0}$ (\AA) & 7.254 & 7.323 (0.95\%) & 4.976 & 5.038 (1.24\%) & 5.011 & 5.173 (3.23\%) & 8.338 & 8.464 (1.51\%) \\
$c{0}$ (\AA) & - & - & 3.950 & 4.038 (2.23\%) & 24.652 & 24.022 (-2.55\%) & 8.111 & 8.240 (1.59\%) \\
$C{11}$ (GPa) & 136.79 & 130.80 (-4.38\%) & 179.79 & 171.42 (-4.66\%) & 160.66 & 148.6 (-7.51\%) & 225.31 & 245.54 (8.98\%)\\
$C_{12}$ (GPa) & 94.74 & 112.81 (19.05\%) & 118.46 & 115.62 (-2.39\%) & 118.85 & 113.17 (-4.78\%) & 84.49 & 131.19 (55.26\%)\\
$C_{13}$ (GPa) & 94.74 & 112.81 (19.05\%) & 111.03 & 117.23 (5.58\%) & 94.28 & 107.23 (13.75\%) & 76.82 & 106.01 (37.98\%)\\
$C_{33}$ (GPa) & 136.79 & 130.80 (-4.38\%) & 256.24 & 209.07 (-18.39\%) & 191.24 & 177.00 (-7.45\%) & 226.17 & 236.62 (4.62\%)\\
$C_{44}$ (GPa) & 24.19 & 30.29 (25.22\%) & 46.23 & 49.32 (6.68\%) & 32.08 & 30.78 (-4.05\%) & 57.04 & 32.43 (-43.13\%)\\
$B$ (GPa) & 108.76 & 118.81 (9.24\%) & 142.65 & 138.39 (-2.99\%) & 125.25 & 125.20 (-0.04\%) & 128.26 & 156.82 (22.26\%)\\
$G$ (GPa) & 22.87 & 18.66 (-18.41\%) & 41.46 & 37.51 (-9.52\%) & 29.31 & 24.71 (-15.71\%) & 65.58 & 47.58 (-27.45\%)\\
$E$ (GPa) & 64.11 & 53.21 (-16.98\%) & 113.39 & 103.21 (-8.97\%) & 81.56 & 69.56 (-14.70\%) & 168.10 & 129.64 (-22.84\%)\\
$\nu$ & 0.402 & 0.425 (5.72\%) & 0.368 & 0.376 (2.17\%) & 0.391 & 0.407 (4.09\%) & 0.282 & 0.362 (28.37\%)\\
\addlinespace[0.1cm]
\hline\hline
\end{tabular}
\end{table*}

\begin{figure*}
    \centering
    \includegraphics[width=0.5\linewidth]{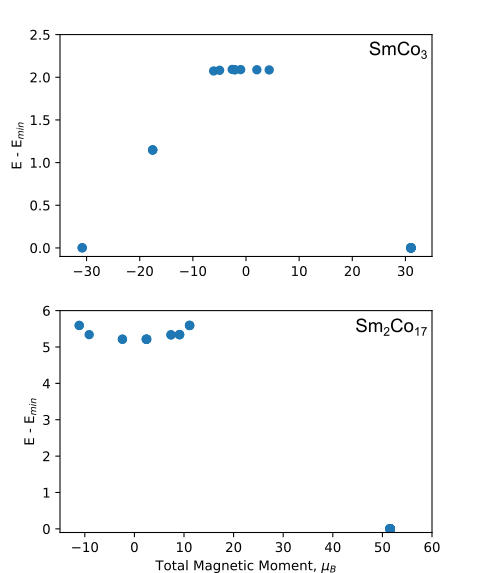}
    \caption{DFT energy comparisons of all stable magnetic configurations in SmCo$_3$ and Sm$_2$Co$_{17}$ were conducted by varying the spin directions across the Wyckoff positions in unit cells. Since only one stable magnetic configuration was identified for SmCo$_2$, it was excluded from the plot.}
    \label{figS1}
\end{figure*}

\begin{figure*}[hbt!]
\centering
\includegraphics[width=\textwidth]{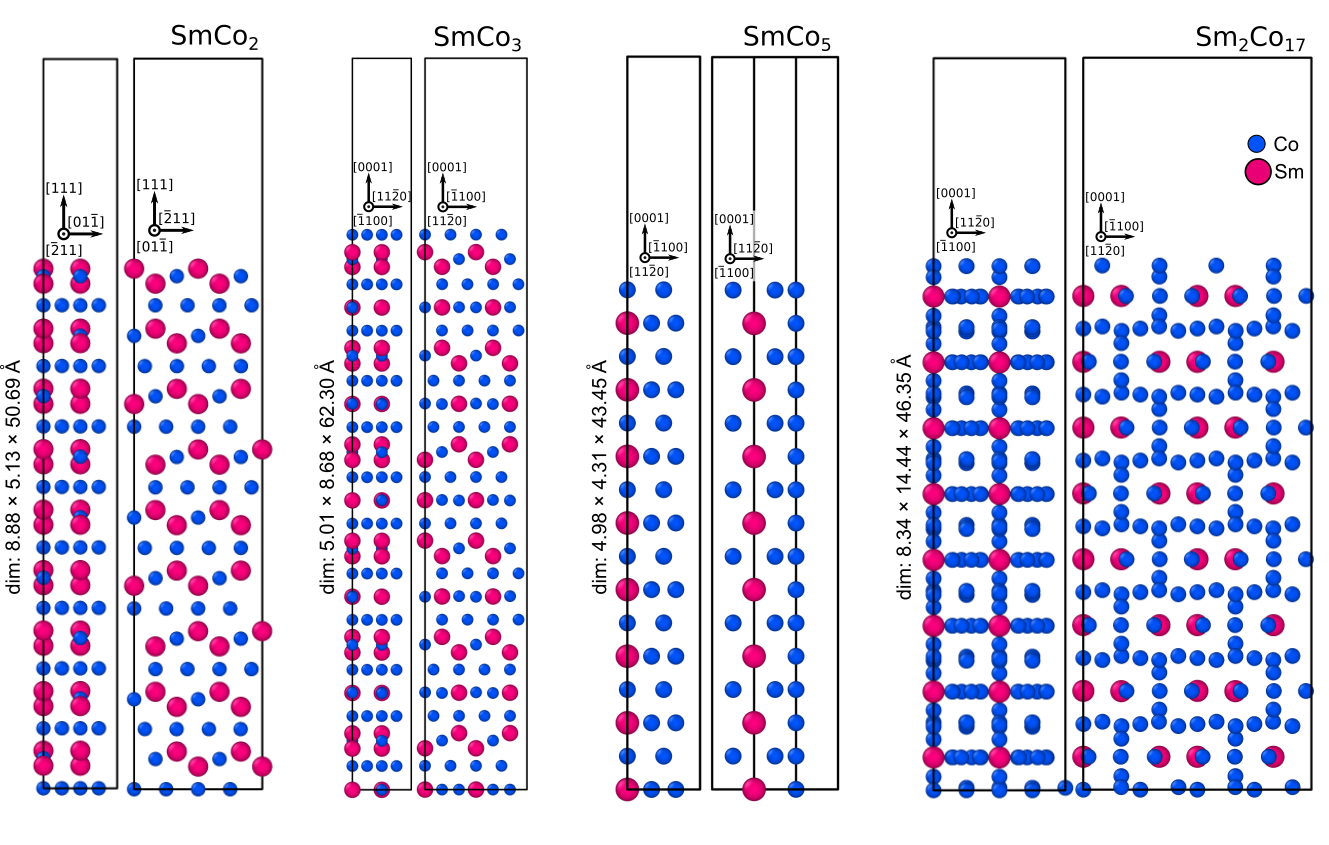}
\caption{The DFT setup for the GSFE calculations of SmCo$_2$, SmCo$_3$, SmCo$_5$, and Sm$_2$Co$_{17}$. The dimensions along the x$\times$y$\times$z axes are annotated next to each structure.}
\label{figS2}
\end{figure*}

\begin{figure*}
    \centering
    \includegraphics[width=0.5\linewidth]{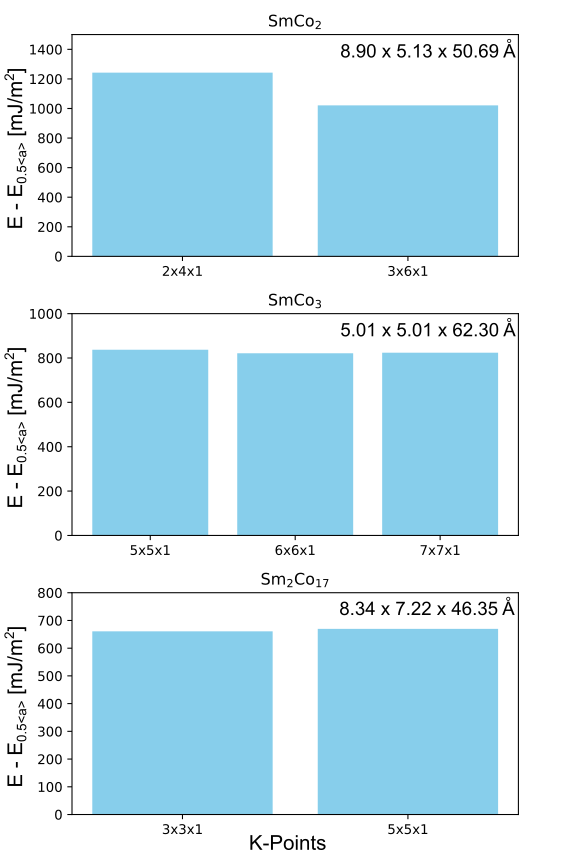}
    \caption{DFT $k$-point benchmarking was performed for SmCo$_2$, SmCo$_3$, and Sm$_2$Co$_{17}$ supercells at states located halfway along the GSFE lines. Due to the differing y-axis lengths of the orthogonalized simulation cells used in the GSFE calculations compared to the conventional unit cell, the final $k$-point meshes were adjusted accordingly, as mentioned in the method section.}
    \label{figS3}
\end{figure*}

\begin{figure*}[hbt!]
\centering
\includegraphics[width=\textwidth]{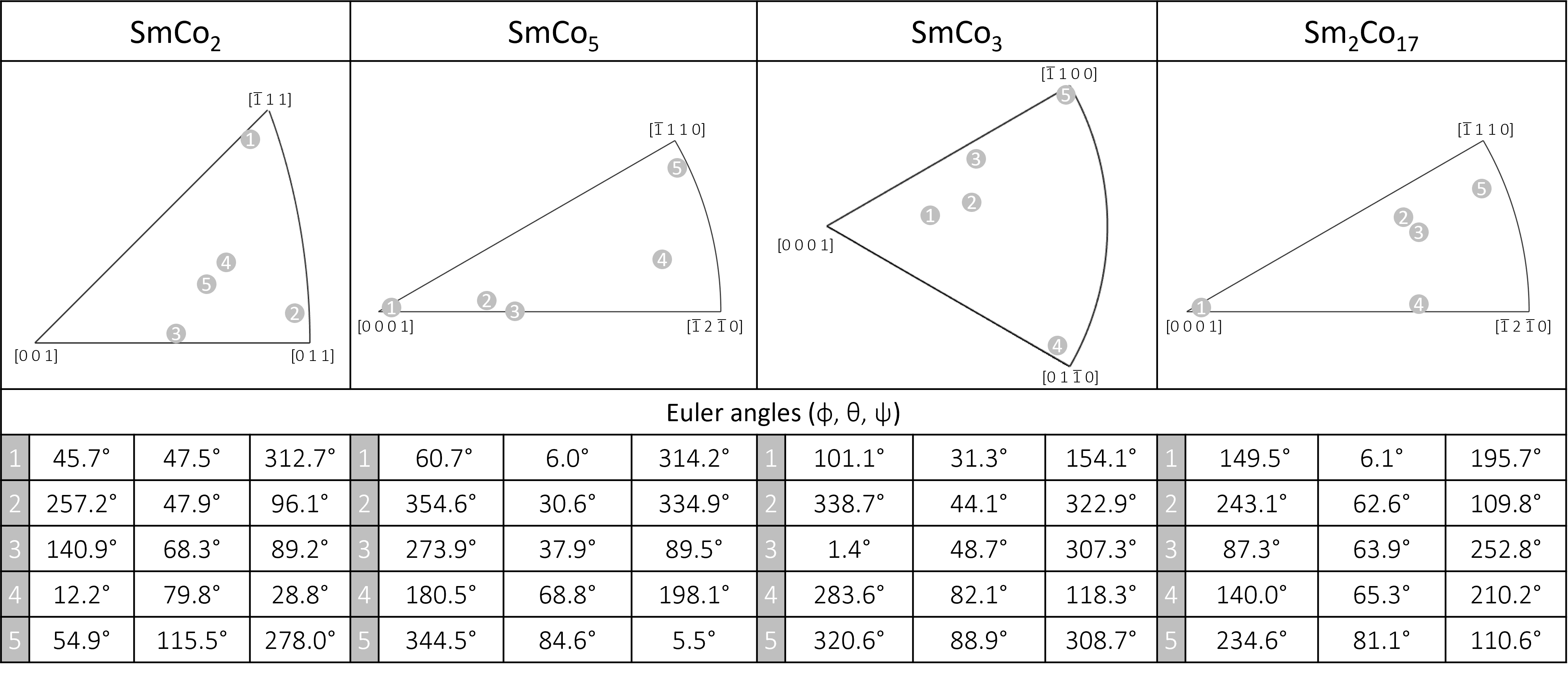}
\caption{Grain orientations for nanoindentation tests in the four investigated Sm-Co phases with their location in the respective standard triangle and the orientations in Euler angles ($\phi$, $\theta$, $\psi$) below them.}
\label{figS4}
\end{figure*}

\begin{figure*}[hbt!]
\centering
\includegraphics[width=\textwidth]{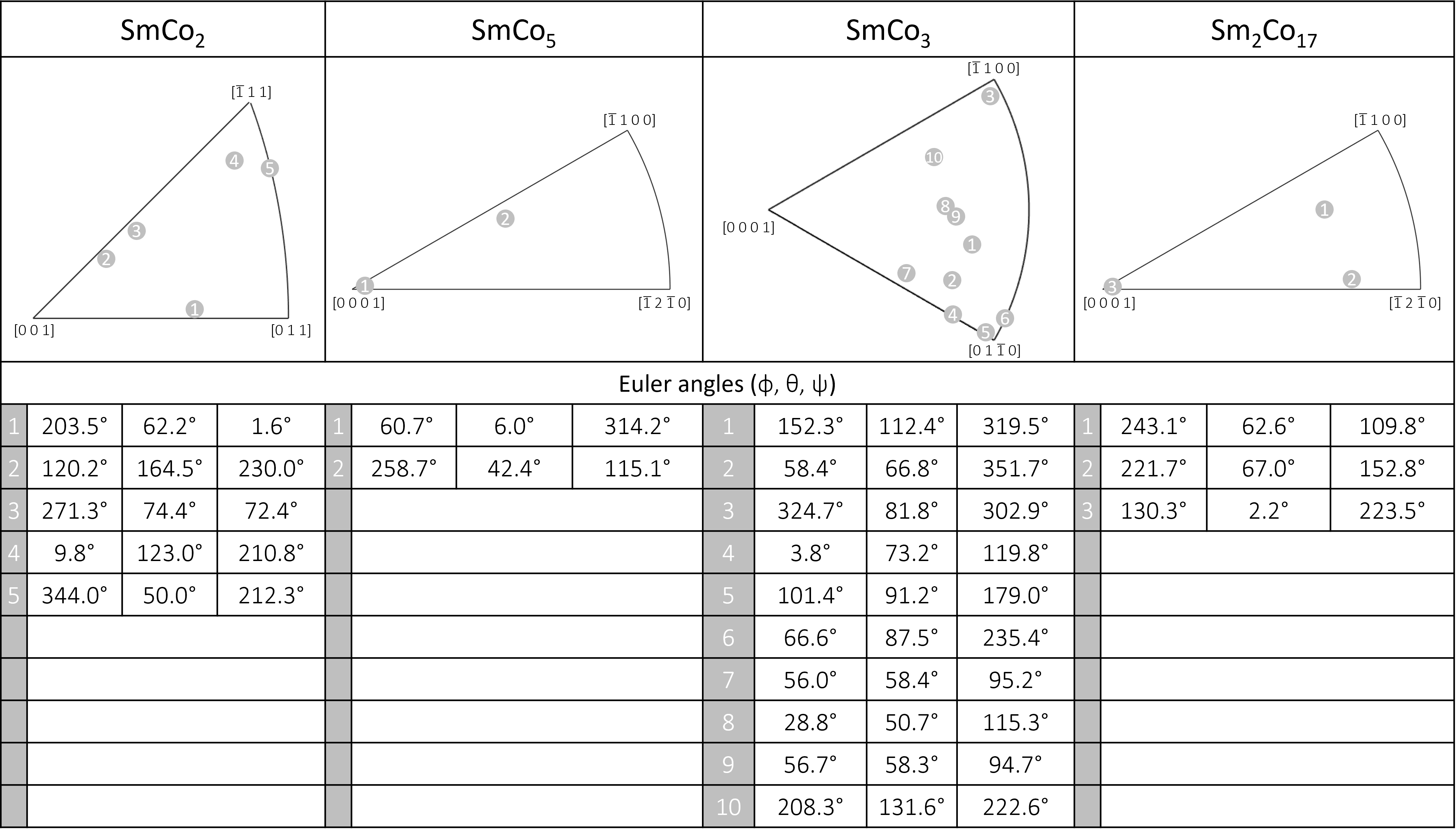}
\caption{Grain orientations for micropillar compression tests in the four investigated Sm-Co phases with their location in the respective standard triangle and the orientation in Euler angles ($\phi$, $\theta$, $\psi$) below them.}
\label{figS5}
\end{figure*}

\begin{figure*}[hbt!]
\centering
\includegraphics[width=\textwidth]{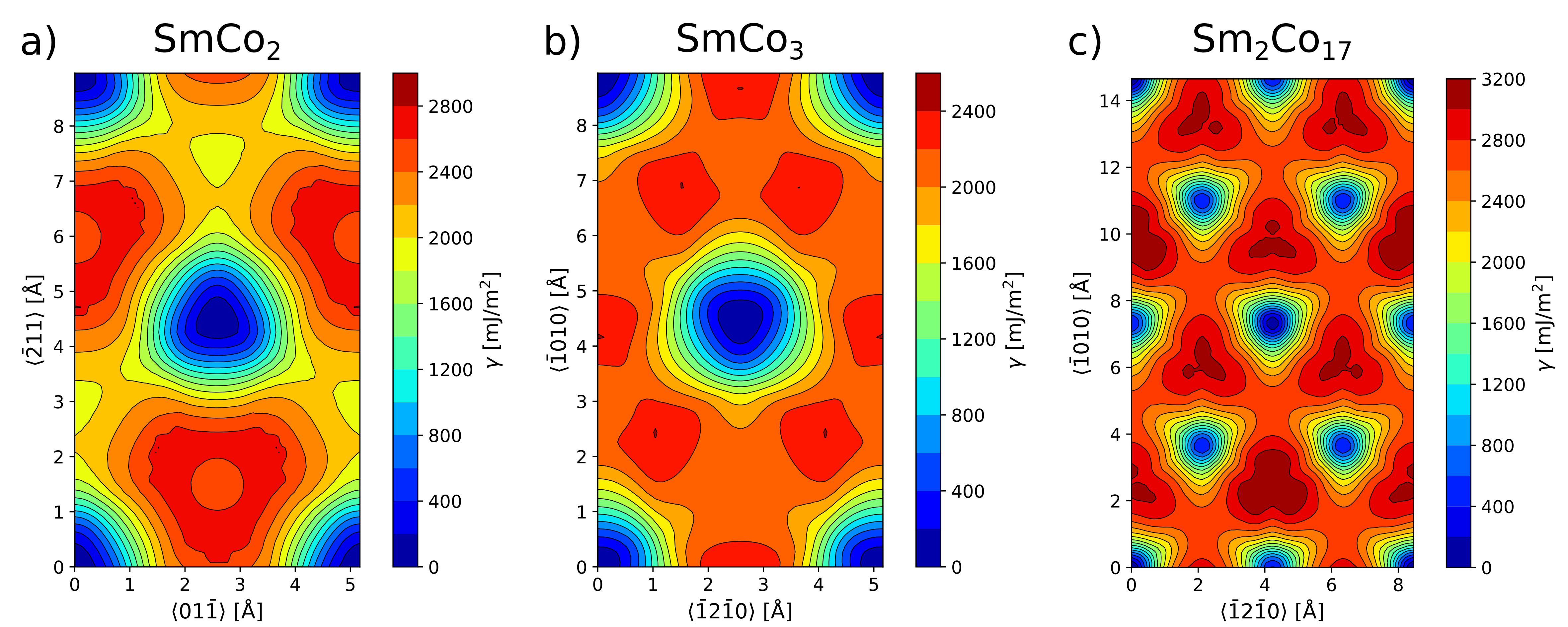}
\caption{Generalized stacking fault energy ($\gamma$) surfaces of a) interlayer II in  SmCo$_{2}$, b) interlayer III in SmCo$_{3}$ and c) interlayer III in Sm$_{2}$Co$_{17}$ calculated using the EAM potential.}
\label{figS6}
\end{figure*}

\begin{figure*}[hbt!]
\centering
\includegraphics[width=0.6\textwidth]{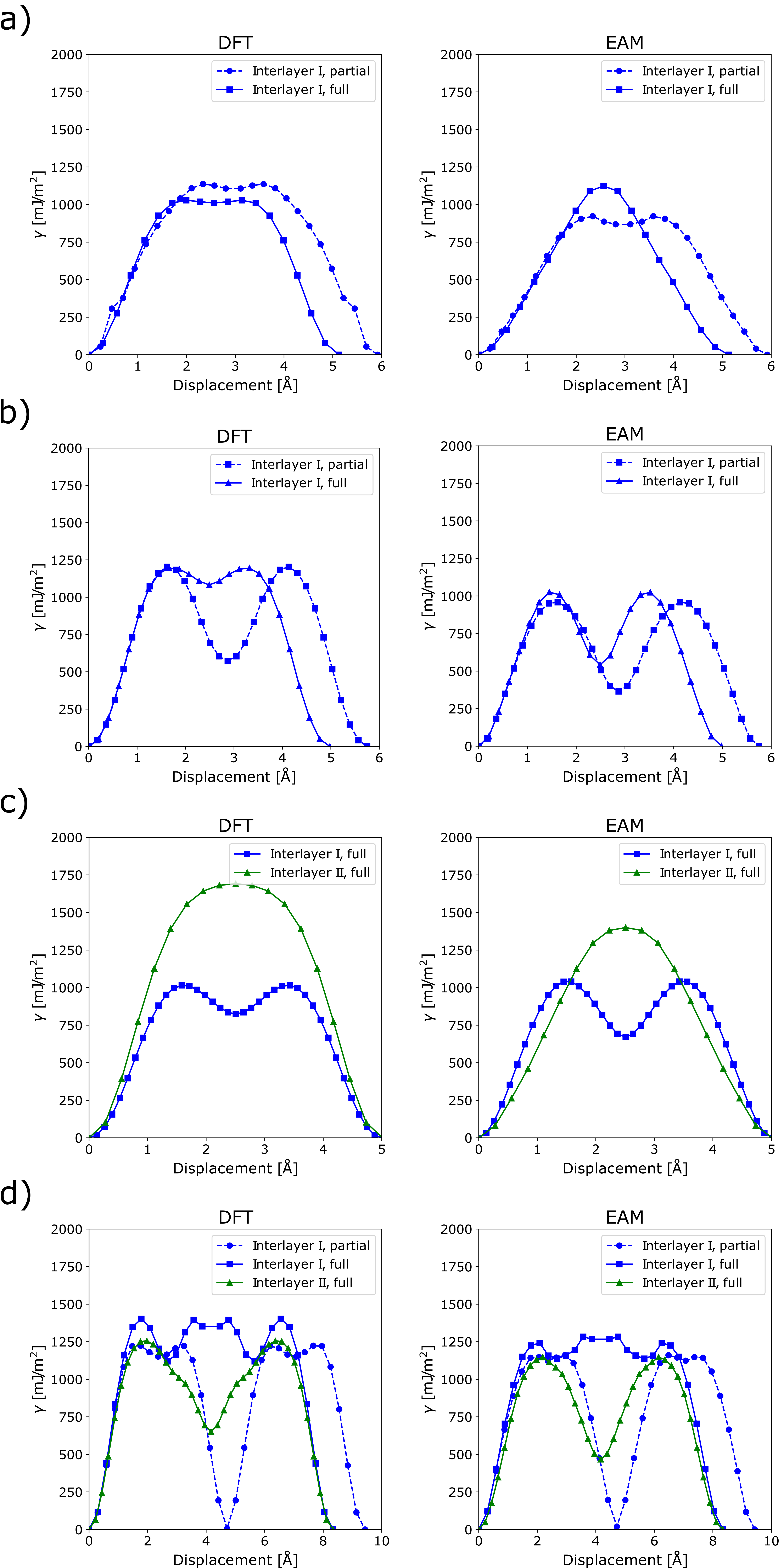}
\caption{Generalized stacking fault energy ($\gamma$) lines in a) SmCo$_{2}$, b) SmCo$_{5}$, c) SmCo$_{3}$, and d) Sm$_{2}$Co$_{17}$ calculated using DFT and reevaluated using the EAM potential.}
\label{figS7}
\end{figure*}

\begin{figure*}[hbt!]
\centering
\includegraphics[width=0.75\textwidth]{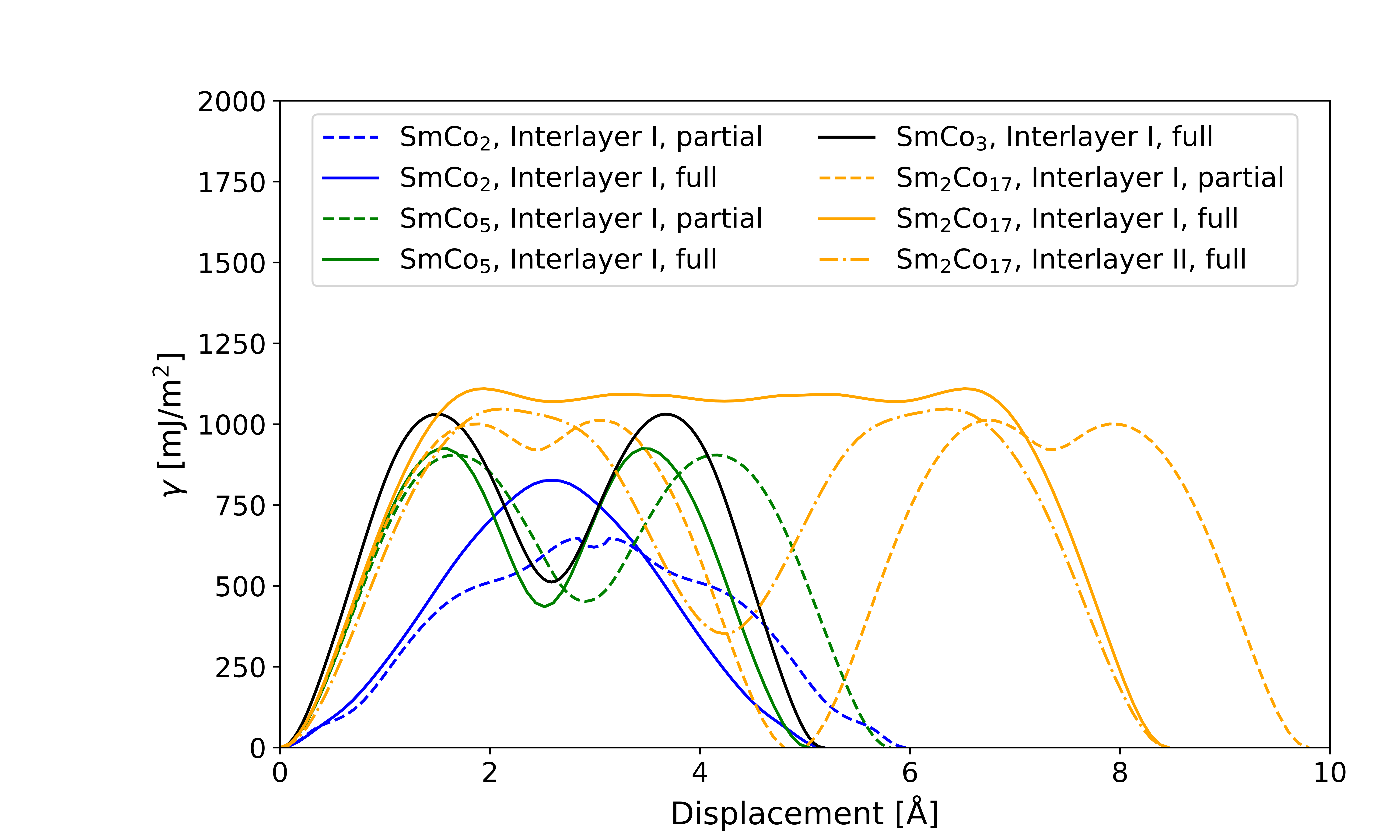}
\caption{Generalized stacking fault energy ($\gamma$) lines along the energetically favorable slip planes and directions in Sm--Co intermetallics calculated using the EAM potential.}
\label{figS8}
\end{figure*}

\begin{figure*}[hbt!]
\centering
\includegraphics[width=\textwidth]{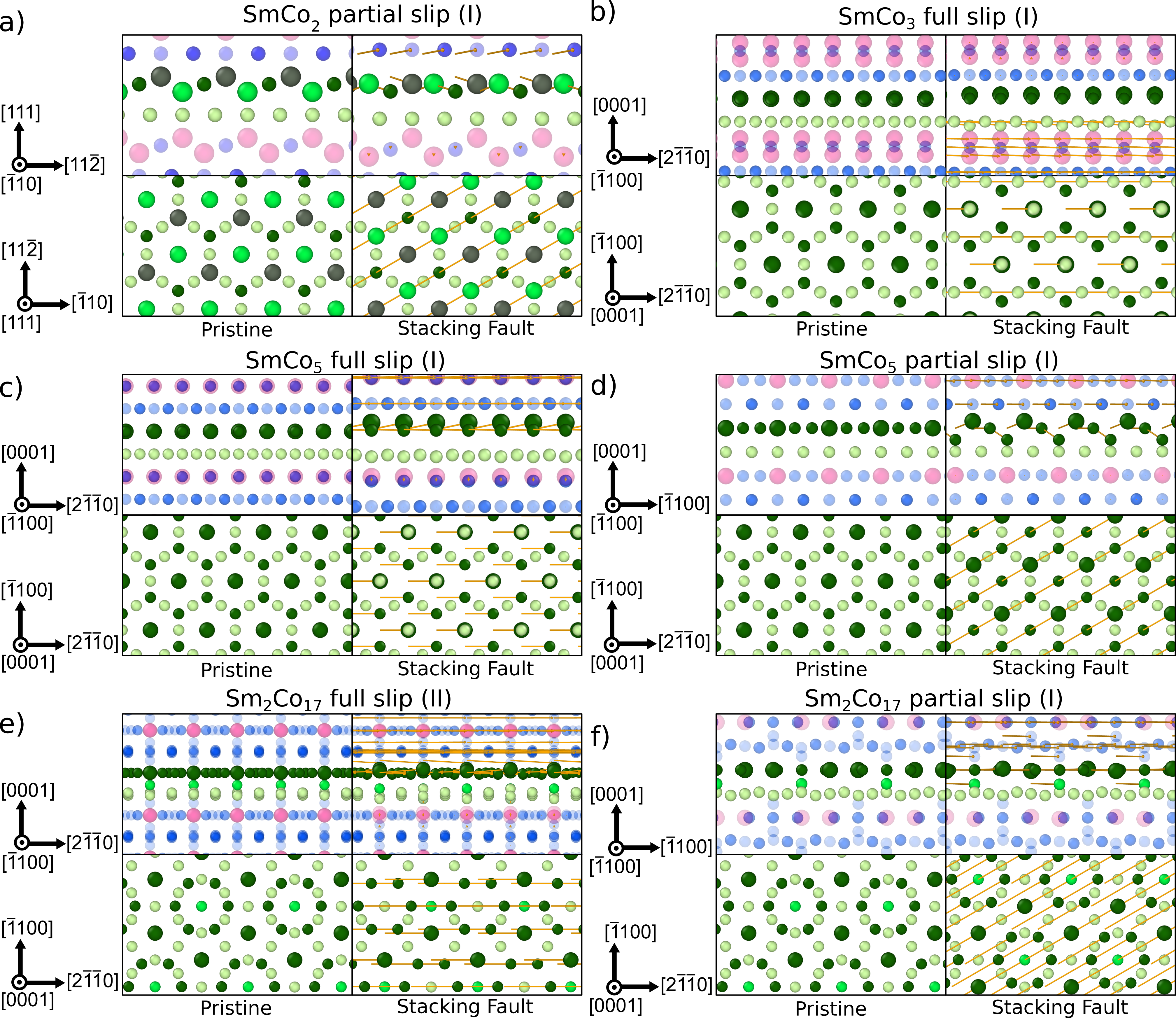}
\caption{Basal and $\{1\,1\,1\}$ stacking fault states in the simulated Sm-Co intermetallics. a) $[1\,1\,1]$ stacking fault after the partial slip $\frac{1}{6}[2\,\bar{1}\,\bar{1}]$ at interlayer I in SmCo$_{2}$. b) $(0001)$ stacking fault halfway along the full slip direction $\frac{1}{6}[2\,\bar{1}\,\bar{1}\,0]$ at interlayer I in SmCo$_{3}$. c) $(0\,0\,0\,1)$ stacking fault after the partial slip $\frac{1}{3}[1\,0\,\bar{1}\,0]$ at interlayer I in SmCo$_{5}$. d) $(0\,0\,0\,1)$ stacking fault halfway along the full slip direction $\frac{1}{6}[2\,\bar{1}\,\bar{1}\,0]$ at interlayer I in SmCo$_{5}$. e) $(0\,0\,0\,1)$ stacking fault halfway along the full slip direction $\frac{1}{6}[2\,\bar{1}\,\bar{1}\,0]$ at interlayer II in Sm$_{2}$Co$_{17}$. f) $(0\,0\,0\,1)$ stacking fault after the partial slip $\frac{1}{3}[1\,0\,\bar{1}\,0]$ at interlayer I in Sm$_{2}$Co$_{17}$. Large and small atoms are Sm and Co atoms, respectively. The atomic layers adjacent to the slip plane are shown in different shades of green, while other atoms are shown as semi-transparent. Only the atoms adjacent to the slip plane are displayed in the $[1\,1\,1]$ and $[0\,0\,0\,1]$ view directions. The displacement vectors of atoms relative to the pristine state are colored in orange.}
\label{figS9}
\end{figure*}

\begin{table*}[!hpt]
\centering
\caption[]{\label{tab2}Summary of the $\{1\,1\,1\}$ and basal slip systems and yield strength ($\sigma_y$) during micropillar compression in the four Sm-Co phases to determine the critical resolved shear stresses (CRSS). The energy barriers of the EAM-relaxed GSFE profiles for the four Sm-Co phases are also summarized. A mean value is provided for the Schmid factor in SmCo$_3$ since pillars in different orientations were considered. The values for basal slip in SmCo$_5$ are marked with an asterisk, indicating that they were published in an earlier work by Stollenwerk et al. \cite{Stollenwerk2024}.}
\centering
\scriptsize
\begin{tabular}{p{0.1\textwidth}p{0.1\textwidth}p{0.1\textwidth}p{0.1\textwidth}p{0.1\textwidth}p{0.2\textwidth}}
\hline\hline
\addlinespace[0.1cm]
Sample & Slip system & Schmid factor & $\sigma_y$ [GPa] & CRSS [GPa] & Energy barrier [mJ/m$^2$] \\
\hline
\addlinespace[0.1cm]
SmCo$_2$ & $(1\,\bar{1}\,1)\langle1\,\bar{1}\,\bar{0}\rangle$& 0.384 & 3.64 ± 0.14 & 1.40 ± 0.05 & 648-826\\
SmCo$_5$ & $(0\,0\,0\,1)\langle2\,\bar{1}\,\bar{1}\,0\rangle$& 0.451 & 3.81* & 1.72* & 905-924\\
SmCo$_3$ & $(0\,0\,0\,1)\langle2\,\bar{1}\,\bar{1}\,0\rangle$& 0.464 ± 0.021 & 4.06 ± 0.17 & 1.88 ± 0.16 & 1031\\
Sm$_2$Co$_{17}$ & $(0\,0\,0\,1)\langle2\,\bar{1}\,\bar{1}\,0\rangle$& 0.384 & 5.40 ± 0.23 & 2.07 ± 0.09 & 1012-1110\\
\addlinespace[0.1cm]
\hline\hline
\end{tabular}
\end{table*}


\end{document}